\documentclass[
 reprint,=
 amsmath,amssymb,
 aps,unsortedaddress]{revtex4-2}
\usepackage{xcolor}
\usepackage{graphicx}
\usepackage{dcolumn}
\usepackage{bm}
\usepackage{amsmath}

\newcommand{\be}{\begin{equation}}
\newcommand{\ee}{\end{equation}}

\begin{document}

\title{Fractonic plaquette-dimer liquid beyond renormalization}

\author{Yizhi You}
\affiliation{Princeton Center for Theoretical Science, Princeton University, 
NJ, 08544, USA}

\author{Roderich Moessner}
\affiliation{ Max Planck Institute for the Physics of Complex Systems, 01187 Dresden, Germany}

\date{\today}
\begin{abstract}
We consider close-packed tiling models of geometric objects---a mixture of  hardcore dimers and plaquettes---as a generalisation of the familiar dimer models. 
Specifically, on an anisotropic cubic lattice, we demand that each site be covered by either a dimer on a $z$-link or a plaquette in the $x-y$ plane. The space of such fully packed tilings has an extensive degeneracy. This maps onto a fracton-type `higher-rank electrostatics', which can exhibit a plaquette-dimer liquid and an ordered phase. We analyse this theory in detail, using height representations and T-duality to demonstrate that the concomitant phase transition occurs due to the proliferation of {\it dipoles} formed by defect pairs. The resultant critical theory can be considered as a fracton version of the Kosterlitz–Thouless transition. A significant new element is its \textit{UV-IR mixing}, where the low energy behavior of the liquid phase and the transition out of it is dominated by local (short-wavelength) fluctuations, rendering the critical phenomenon beyond the renormalization group paradigm. 
\end{abstract}

\maketitle

\section{Introduction}
Close-packed tiling problems provide a fertile platform to demonstrate how many-body systems with constraints give rise to a rich set of phenomena. 
A prototypical example of this class is the close-packed dimer model\cite{kasteleyn1961statistics,kasteleyn1963dimer,fisher1963statistical,rokhsar1988superconductivity,henley1997relaxation,henley2004classical,moessner2011quantum} on a wide variety of lattices, whose low-energy configuration space contains one dimer attached to each site. This constraint engenders an extensive number of configurations with the same internal energy, so the equilibrium behavior is controlled by entropic contributions to the free energy, rather than energetic competition. Notwithstanding the simplicity of the classical dimer model, these constrained models provide illuminating insight into a wide range of physical situations, including resonating valence bond  liquids\cite{moessner2001resonating}, classical spin ice\cite{castelnovo2008magnetic}, order by disorder and unconventional phase transitions\cite{1986JSP....44..729Y} etc. 

The close-packed dimer models play a central role in statistical physics due to their relationship to 2D height models\cite{1997PhRvB..5514935Z,2005PhRvL..94w5702A} and 3D gauge theories\cite{2003PhRvL..91p7004H}. In particular, a step towards the understanding of classical dimers on a 2D bipartite lattice lies in the fact that the close-packed constraint of the dimer can be represented by a fluctuating scalar field so the entropy fluctuation can be mapped to a statistical height models\cite{kasteleyn1961statistics,fisher1963statistical,1984JPhA...17.3559N}. Following the same spirit, the close-packed dimers on the 3D cubic lattice can be expressed in terms of a magnetic field with monopole-free conditions, and the entropy of its fluctuations is captured by a 3D classical gauge theory. If we further take quantum effects into consideration, the quantum resonance of dimers can generate a richer class of quantum phases.
Depending on lattice structure and dimensionality, at long wave-lengths the gauge field dynamics can either exhibit confinement or else can be described by a discrete or continuous gauge structure, respectively, characterized by the framework of $Z_2$ and U(1) spin liquids\cite{rokhsar1988superconductivity,moessner2001resonating,moessner2001short,2003PhRvB..68r4512M,hermele2004pyrochlore}.

Close-packed tiling models provide an exquisite framework within which to study such emergent phenomena. This line of research has been extended into several directions, including close-packed dimers on quasi-crystals\cite{flicker2020classical}, close-packed plaquette and cube (and also mixed\cite{2015PhRvL.114s0601R}) models\cite{xu2008resonating,pankov2007resonating,2019PhRvE..99e2129V} on the square and cubic lattices\cite{you2019emergent}. While most close-packed tiling problems share similar behavior, including extensive(or subextensive) ground state degeneracy and zero-temperature entropy, the plaquette tiling system contains additional peculiar features as the monomer defects (unpaired sites) display restricted motion. In particular, a single monomer in the close-packed plaquette system cannot move alone, while a pair of monopoles between links can only move along the transverse direction. Such restricted motion is
reminiscent of \textit{fracton phenomenon}-- a quasiparticle with restricted motion\cite{Haah2011-ny,vijay2017generalization,Chamon2005-fc,pretko2017generalized,pretko2017subdimensional,vijay2016fracton,yan2019rank}. In Ref.~\cite{xu2008resonating,you2019emergent}, it was demonstrated that the plaquette tiling constraint could be interpreted as a higher-rank gauge theory\cite{xu2007bond,rasmussen2018intrinsically,prem2018pinch,gromov2020fracton,nandkishore2021spectroscopic} whose charge is conserved on a subdimensional manifold.

In this paper, we try to extend the close-packed tiling models into a broader class of geometric configurations. To begin with, we consider a close-packed dimer-plaquette tiling problem on a cubic lattice. The configuration (`Hilbert' space) consists of all the plaquette and dimer patterns obeying
the following constraint: every site in the cubic lattice is connected to either a dimer on the $z$-link or a plaquette on the $xy$ plane. Around each site, there are thus six possible configurations, four with plaquettes living at the four adjacent squares on the $xy$ plane, or two with dimers living on the $z$-link above or beneath the site as Fig.~\ref{fig1}. The space of such fully-packed tiling patterns has an extensive degeneracy with large entropy. 

The free energy landscape is generated by the  fluctuations between distinct close-packed configurations. 
We analyse this close-packed dimer-plaquette model by mapping it into a `higher-rank electrostatics' whose charge is conserved on each $xz$ and $yz$ plane\cite{shirley2018fractional,shirley2018foliated}. We in turn solve this `higher-rank electrostatics' by mapping it to a height model with subsystem U(1) symmetry\cite{you2020higher,seiberg2020exotic,xu2007bond,paramekanti2002ring,karch2020reduced,gorantla2021modified} and demonstrate that such theories can support a liquid phase with power-law decaying correlation functions. 
By tuning a stiffness, we observe an ordering transition, for which we provide a complete discussion from the perspective of T-dualiy, where a low-temperature height model can be mapped into a high-temperature classical rotor model in 3D with subsystem U(1) symmetry. The phase transition is driven by the proliferation of a set of dipoles formed by defect pairs. 

The most striking feature of these close-packed dimer-plaquette models is their \textit{UV-IR mixing}, where the low energy effective theory is controlled and dominated by discontinuity field patterns\cite{seiberg2020exotic}. As a result, the liquid phase and the ordering transition is beyond the renormalization group paradigm as the low-energy behavior at criticality is manipulated by local fluctuation at short wave-lengths. Such UV-IR mixing\cite{gorantla2021modified,xu2007bond,paramekanti2002ring,you2019emergent,you2020fracton,you2021fractonic} yields a new class of critical phenomenon beyond the renormalization perspective. Finally, we  also generalize our discussion to a broader class of close-packed problems with trimer-dimer mixture patterns and fractal dynamics.

\section{Plaquette-dimer coverings of the cubic lattice}\label{sec:one}
\begin{figure}[h]
  \centering
      \includegraphics[width=0.47\textwidth]{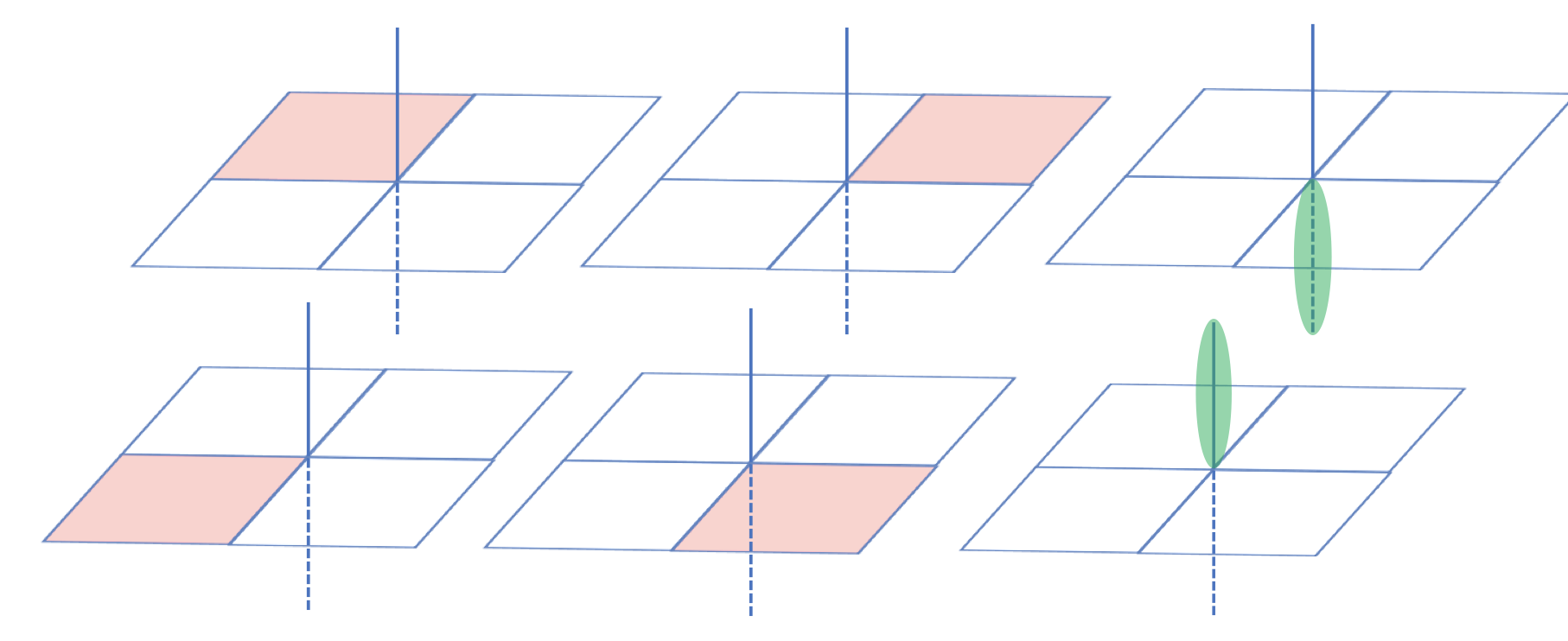}
  \caption{Close-packed tilings of the cubic lattice. Each site is either part of a plaquettes(red) in the x-y plane or a z-dimer(green).} 
\label{fig1}
\end{figure}

To set the stage, we begin with the close-packed tilings of the cubic lattice as Fig.~\ref{fig1}. Each site is either connected to a dimer on an adjacent z-link or to a plaquette on an adjacent x-y square. 
The resulting configuration space of close-packed dimer-plaquette patterns has an extensive degeneracy and non-vanishing entropy at zero temperature. 

To analyse the close-packed patterns, we account for the local dimer-plaquette constraint by encoding the plaquette and dimer coverage as a higher-rank electric field\cite{shirley2018fractional,you2019emergent,gromov2019towards},
\begin{align} 
E_{xy}=\eta P_{xy}, ~E_{z}=\eta D_{z}\ . \label{def}
\end{align}
$E_{xy}$ lives on the center of each x-y plaquette while $E_z$ lives on each z-link.
$P_{xy},  D_{z}$ refers to the number of plaquettes and dimers living on the x-y plaquettes and z-links, respectively. The $\eta$ is the bipartite lattice factor with an alternating sign structure. As we can uniquely associate each x-y plaquette(and z-dimer) with a site at $(x,y,z)$, we can define the bipartite lattice factor as $\eta=(-1)^{x+y+z}$. Based on this notation, the dimer-plaquette constraint can be interpreted as the Gauss-law,
\begin{align} 
\Delta_x \Delta_y E_{xy}+\Delta_z E_{z}=\eta (1-Q)
\label{cons}
\end{align}
Here $\Delta_i$ is the lattice difference and we set the lattice constant $a=1$.
$Q$ denotes the monomer number on the site.
When considering the close-packed configurations we choose $Q=0$, with the staggering background charge $\eta$ indicating each site is either connected with a dimer or plaquette. To satisfy this local constraint, one can parameterize the electric field as,
\begin{align} \label{hei}
E_{xy}=-\Delta_z h+\bar{E}_{xy},E_{z}=\Delta_x\Delta_y h+\bar{E}_{z},
\end{align}
Here $h$ is a discrete integer-valued field living on the dual lattice at the center of each cube which characterizes the local fluctuation of the dimer-plaquette pattern. $\bar{E}_{z},\bar{E}_{xy}$ are background patterns satisfying the constraint in Eq.~\ref{cons} due to the staggered background charge. We can simply take the configuration $\bar{E}_{z}=\eta, \bar{E}_{xy}=0$ for the dimer columnar phase or $\bar{E}_{z}=0, \bar{E}_{xy}=\eta$ for the plaquette columnar phase. It is worth mentioning that a large number of distinct plaquette-dimer configurations can be connected by changing the value of $h$ locally. In the meantime, the background patterns $\bar{E}_{z},\bar{E}_{xy}$ are responsible for the global topological sectors that cannot be connected by local $h$ fluctuations. We will return to this issue in Sec.~\ref{sec:top}.

For the close packing problem, the partition function contains a summation of all the allowed plaquette/dimer configurations with equal Boltzmann weights. Since there are no energetic terms in the partition function, the free energy consists only of the entropy. If we coarse-grained the $E_{xy},E_z$ field, the flippable configurations with $\bar{E}_{xy},\bar{E}_z=0$ corresponds to a larger number of microscopic states, and hence a larger coarse-grained entropy, than the non-flippable configuration $\bar{E}_{xy},\bar{E}_z \neq 0$. 
Indeed, a flippable cube that can resonate between two x-y plaquettes and four z-dimers has zero average $E_{xy}, E_z$, so the coarse-grained energy should effectively favors such flippable patterns. 
This motivates the following ansatz for the height field:
\begin{align} \label{th}
&\mathcal{Z}=\int \mathcal{D}E_z~ \mathcal{D}E_{xy}~ e^{-\beta(E_z^2+E_{xy}^2)}\nonumber\\
&=\int \mathcal{D}h~ e^{-\beta[(\partial_z h)^2+(\partial_x\partial_y h)^2]+\alpha \cos(2\pi h)+...}\ .
\end{align}
Here we replace the lattice difference with differentials in coarse-graining.
The term $\alpha \cos(2\pi h)$ imposes the constraint that $h$ takes discrete integer values. Depending on the relevance of this term, this constraint can be released and the resultant height field becomes a continuous variable.

Bear in mind that the ground state manifold comprises the extensively degenerate patterns with close-packed configurations so the $e^{-\beta((\partial_z h)^2+(\partial_x\partial_y h)^2)}$ describes the entropy of each of the coarse-grained fields, with $\beta$ the resulting stiffness.  When $\beta$ is large, the entropic fluctuations strongly
order. This kind of effect, where fluctuations around ordered state dominate over an ensemble of disordered configurations with lesser fluctuations, is known as order by disorder.  

Long-wavelength quantities such as $\beta$ can in principle be tuned by adding microscopic interactions between dimers and plaquettes. This can thus drive the system across a phase transition, see e.g.~\cite{2005PhRvL..94w5702A} in the case of a dimer model on the square lattice. If we add attractive or repulsive interactions between adjacent plaquettes/dimers, the locally flippable configurations with columnar dimer/plaquette patterns are microscopically favored/suppressed. Whether or not a liquid phase exists in a given model such as ours cannot in general be read off from its microscopic formulation directly; rather, one needs to determine the value of the prefactors/relevance of the various terms in the effective action independently. The effective description does allow an analysis of the novel phases, and the transitions between them, where they arise.

Thus, as we shall see below, when $\beta <\pi/2$, the system is in the liquid phase, with dimer-dimer (and plaquette-plaquette) correlations displaying a power-law decay as Eq.~\ref{co}. The phase transition between the liquid and ordered phase can be characterized by a fractonic Kosterlitz-Thouless transition which we address in detail in Sec.~\ref{sec:dual}.

\begin{figure}[h]
  \centering
      \includegraphics[width=0.25\textwidth]{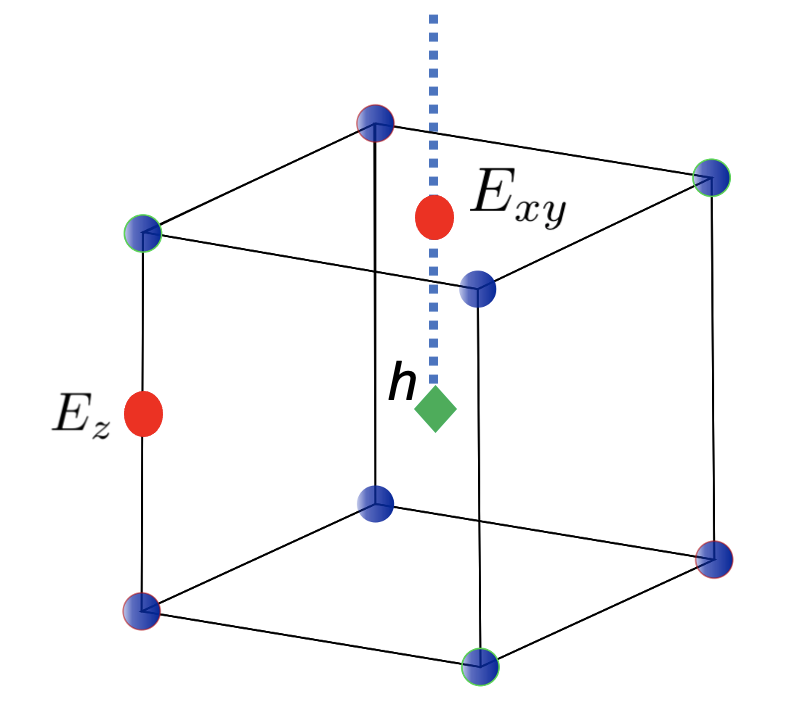}
\caption{The electric fields $E_{xy},E_z$(red dot) live at the center of the x-y plaquette or at the center of the z-link. The height field $h$ (green) lives at the center of each cube on the dual lattice. } 
\label{fig2}
\end{figure}

This theory has a special feature as the `effective Hamiltonian' of the height field (which encodes the entropy) contains a higher-order derivative $\partial_x\partial_y$ so the $k_x,k_y$ axes are dispersionless, hosting  a `sub-extensive number' of zero modes, i.e.\ the zero modes reside on a lower-dimensional manifold. This also implies that the system has an additional subsystem symmetry as $h \rightarrow h+g(x,z)+f(y,z)$. 

We first consider a Gaussian fixed point theory,
\begin{align} \label{fix}
\mathcal{Z}=\int \mathcal{D}h~ e^{-\beta[(\partial_z h)^2+(\partial_x\partial_y h)^2]}
\end{align}
At this point, we ignore the discreteness of $h$, the validity of which will be scrutinized later.
We calculate the Green function of the free height field,
\begin{align} 
\langle h(q)h(q) \rangle =\frac{1}{\beta(q_z^2+q_y^2q_x^2)}\ .\label{green}
\end{align}
As the Green function contains a set of nodes on the $k_x,k_y$ axis, the asymptotic behavior of the height correlator at large $z$ exhibits double logarithmic scaling as,
\begin{align} 
\langle e^{-h(0)h(z)} \rangle =e^{-\frac{1}{4\pi \beta} [\ln(z)]^2}\label{doublel}
\end{align}
The double linear divergence of the energy spectrum at $k_x\rightarrow 0, k_y \rightarrow 0$ leads to the special double logarithmic scaling in the correlator, yielding a decay faster than any power-law. Thus, the height operator is short-ranged correlated along the z-direction. 
In addition, the two-point correlator outside of the $z$-direction is short-ranged,
\begin{align} 
\langle e^{-h(0,x,y)h(0,0,0)} \rangle \rightarrow 0
\end{align}
This is manifested by the additional subsystem symmetry $h \rightarrow h+g(x,z)+f(y,z)$.  Thus, we can conclude that the two-point height correlator is always short-ranged, so the  $\cos{2\pi h}$ operator is irrelevant for any $\beta$. 

An intuitive argument to elucidate the absence of long-range correlation goes as follows. Due to the subsystem symmetry of the classical height field, it can only fluctuate along the $z$ direction so we regard it as a $1D$ subdimensional classical field. For a 1D classical system, there is no long-range order due to its strong fluctuations.
However, this does not imply we can ignore the discreteness of the height field as higher-order operators like $\cos{2\pi h}$, albeit with more derivatives, could be more relevant at large $\beta$. 

We also calculate the dimer and plaquette correlation functions,
\begin{align} 
&\langle E_z(0) E_z(r) \rangle =\langle \partial_x\partial_y h(0) \partial_x\partial_y h(r) \rangle \nonumber\\
&=\frac{1}{\beta}\left(\frac{1}{z^2+x^2y^2}-\frac{2z^2}{(z^2+x^2y^2)^2}\right)
\end{align}
\begin{align} 
&\langle E_{xy}(0) E_{xy}(r) \rangle =\langle \partial_z h(0) \partial_z h(r) \rangle \nonumber\\
&=\frac{1}{\beta}\left(\frac{1}{z^2+x^2y^2}-\frac{2x^2y^2}{(z^2+x^2y^2)^2}\right)\label{co}
\end{align}
Both display power-law decay with strong anisotropy. Based on this observation, the extensive entropy  of such close-packed patterns engenders a liquid phase where the dimer and plaquette patterns strongly fluctuate. Such anisotropic power-law correlation also result in the pinch point singularity in momentum space \cite{prem2018pinch,nandkishore2021spectroscopic,benton2021topological} which reflects the Gauss-law constraint in Eq.~\ref{cons}.

Now we return to the relevance of the $\cos(2\pi h)$ term. As the correlation function $\langle e^{- h(0)h(z)} \rangle =e^{-a [\ln(z)]^2}$ with double logarithmic decay vanihes faster than any power law (but slower than exponentially), and is short-ranged otherwise due to subsystem symmetry, the $\cos(2\pi h)$ term is always irrelevant. 
However, there exist higher-order operators $\cos{(2\pi \nabla_x h)},\cos{(2\pi \nabla_y h)}$ that are also allowed by the integer constraint. These operators correspond to a pair of height fields separated along the x(y)-link on the dual lattice.
\begin{align} 
&\langle e^{-4\pi^2 ( h(0) h(e_x) h(0,y,z)h(e_x,y,z))} \rangle\nonumber\\
&=\langle e^{-(2\pi \nabla_x h(0)  ~2\pi\nabla_x h(0,y,z))} \rangle
=\frac{1}{(z^2+y^2)^{\frac{\pi}{\beta}}},\nonumber\\
&\langle e^{-4\pi^2( h(0) h(e_y) h(x,0,z)h(x,e_y,z))} \rangle\nonumber\\
&=\langle e^{-(2\pi\nabla_y h(0)  ~2\pi\nabla_y h(x,0,z))} \rangle
=\frac{1}{(z^2+x^2)^{\frac{\pi}{\beta}}}
\label{dipcor}
\end{align}
Taking the lattice spacing $a$ to be fixed at unity, the four-point correlator on the thin stripe becomes the correlation between `dipoles'. Notice that the dipole correlation function $e^{-[2\pi \nabla_y h(0)  2\pi\nabla_y h(x,0,z)]} $ is only nonzero when they are at the same $xz$ plane transverse to the dipole orientation. The dipoles effectively behave like a $2D$ classical field with restricted motion and algebraic correlations within each slab. When $\beta > \pi/2$,  the $\cos(2\pi \nabla_x h),\cos(2\pi \nabla_y h)$ term becomes relevant and hence drives the liquid phase toward an ordered phase with crystalline
patterns (dimer or plaquette ordered)\cite{you2021fractonic,you2019emergent,paramekanti2002ring,xu2007bond}. Our current scaling argument only implies the existence of a phase transition between the liquid and the ordered phase.
The microscopic pattern of the crystalline phase will be determined by the concrete Hamiltonian, which can vary if we tune the interactions between favouring plaquettes or dimers. In addition, there are other symmetry allowed terms including the parallel-jump operator\cite{xu2007bond} $\cos(2\pi \nabla^2_i h)$ with higher-order derivatives. Due to the subextensive number of low energy modes at high momentum, these operators representing 'rough patterns with local fluctuations' cannot be ignored. In Ref.~\cite{xu2007bond}, it was shown that the parallel-jump operator are less relevant for integer fillings.

In conventional critical phenomena within the RG scheme, an operator's correlation function either decays as a power-law function or exponentially. Such scaling implies that a higher-order derivative of the operator has a faster decay with a smaller scaling dimension and hence is less relevant. However, the double logarithmic scaling in the two-point correlation function in Eq.~\ref{doublel} contains a peculiar feature that its higher-order derivative operator turns out to be more relevant and longer-ranged. Thus, the critical point is driven by dipole proliferation. Actually, the double logarithmic scaling in Eq.~\ref{doublel} is a necessary feature of UV-IR mixing where the critical point is controlled by short wave-length physics. We will return to this point in Sec.~\ref{sec:uv}.

It is noteworthy that determining the relevance and the scaling dimension of the dipole operator is considered based on the subsystem slab instead of the entire space-time dimension. General renormalization group reasoning implies that classical operators are irrelevant when the associated scaling dimension exceeds the space dimension. However, due to the constrained form of the dipole correlators, which only exhibit a power-law decay on a reduced spatial region, i.e.,  on the 2D slab, one expects that the condition for irrelevance should be modified, and the instability should appear only provided the associated scaling dimension exceeds that of the subsystem, the 2D plane. Thus, when evaluating the scaling dimension of the vertex operator $ \cos(2\pi \nabla_x h)$, we only consider its scaling within the y-z plane. Renormalization on the subsystem is a significant new element in our critical theory, and it implies that the IR theory allows discontinuous field configurations due to the subsystem symmetry. Such discontinuous field configurations in the effective field theory are manifested by the lines of nodes on the $k_x,k_y$ axis in the height field Green function in Eq.~\ref{green}.

When $\beta <\pi/2$, the dimer-plaquette liquid phase resembles a classical dimer `liquid' on a 2D bipartite lattice where the fluctuations produce power-law decaying dimer correlations. However, the plaquette-dimer liquid we study here is intrinsically distinct from the classical dimer liquid in the following sense. While the constraint of the plaquette-dimer liquid is engendered by the generalized Guass law in Eq.~\ref{cons}, the monomers ($Q$ charges) obey a special subsystem charge conservation law,
\begin{align} 
&\int dx dz~(\Delta_x \Delta_y E_{xy}+\Delta_z E_{z})=\int d x dz~ Q=0, \nonumber\\
&\int dy dz~(\Delta_x \Delta_y E_{xy}+\Delta_z E_{z})=\int dy dz~ Q=0,
\end{align}
If we impose periodic boundary conditions, the monomer charges along each x-z and y-z plane are conserved. Thus, a single monomer excitation that corresponds to an unpaired site with no dimer and plaquette adjacent to it can only move along the z-direction by exchanging its position with the z-dimers along the path. In the meantime, the monomer alone cannot move on the x-y plane. Instead, only a pair of monomers (i.e. a dipole) separated along an x(y) link can move together along the transverse direction by exchanging its position with the plaquette in the x-y plane. 

Such monomer excitations with restricted motions are known as \textit{fractons}. A fracton is a type of quasiparticle with restricted mobility that was first introduced in the context of exactly solvable spin liquid models \cite{Haah2011-ny,Chamon2005-fc,vijay2016fracton}. Theoretically, the characterization of fractons is framed by the language of higher-rank gauge theories, which encode the immobility of fractons in a set of subsystem charge conservation conservation laws\cite{pretko2017generalized,gromov2019towards}.

\section{UV-IR mixing: What it is and how it appears?}\label{sec:uv}
 Up to this stage, we have demonstrated the possibility of a close-packed plaquette-dimer liquid with power-law decaying correlation functions. By tuning the stiffness $\beta$, there exists a phase transition between the plaquette-dimer liquid and an ordered phase. While such an ordering is known from classical dimers on various crystalline structures, our plaquette-dimer liquid transition has a unique feature denoted "UV-IR mixing", where the short wavelength physics plays a crucial rule. As we elucidated in our previous discussion Sec.~\ref{sec:one}, the transition toward an ordered phase is triggered by the dipole operator $\cos(2\pi \partial_x h),\cos(2\pi \partial_y h)$ while the operator $\cos(2\pi h)$ itself remains irrelevant. This anomalous scaling is peculiar and counter-intuitive as a renormalization group analysis or dimension counting generally implies that higher-order operators should be less relevant.

The divergence of the correlation length in the critical region implies the effective interaction at IR becomes long-ranged and hence requires us to visualize the system at larger scales. A complete understanding of critical phenomena has famously been accomplished via the development of the renormalization group\cite{landau,wilson1983renormalization,fisher1983scaling,fisher1998renormalization}. In particular, the universal properties of a wide class of random or statistical systems can be understood by coarse-graining: integrating out the short wave-length modes and focusing on the resulting (renormalisation flow of the) long wave-length behavior. Based on this observation, the critical phenomenon exhibits many universal properties that are independent of the UV Hamiltonian but only reflect symmetry and dimensionality. For instance,  the correlation function at the critical point has a universal power-law exponent $\langle C(r) C(0) \rangle=\frac{1}{r^{D-2+\eta}}$ with $D$ being the space dimension and $\eta$ the anomalous dimension correction. The field pattern at low energy, which controls the IR behavior, is thus determined merely by the spatial dimension and an exponent that are insensitive to UV cut-offs\cite{xu2007bond,seiberg2020exotic,you2021fractonic}.

However, the plaquette-dimer liquid phase we discuss here is a peculiar example that escapes this renormalization group picture. The power-law correlation of the dimer/plaquette liquid is spatially anisotropic with only a $C_4$ symmetry, and its exponent does not match any known universality. In addition, the operator $\cos{h}$ always remains irrelevant and short-range correlated while the
the dipole operator $\cos(2\pi \nabla_x h) (\cos(2\pi \nabla_y h))$ exhibits algebraic correlation if and only if evaluated within the same y-z (x-z) plane. Such a dipole correlator can also be understood as the  correlation between {\it four} height fields at the corners of a think stripe. It becomes relevant when $\beta$ grows and hence drives the liquid towards the ordering transition.

In this spirit, we can measure the asymptotic behavior of the four-point height correlator defined at the corners of a large rectangle that respects subsystem U(1) symmetry,
\begin{align} \label{fourp}
&\langle e^{-[ h(0) h(x) h(0,0,y)h(0,x,y)]} \rangle =e^{-a [\ln(r_x) \ln(r_y)]}
\end{align}
The double linear divergence of the energy spectrum as $k_x\rightarrow 0, k_y \rightarrow 0$ leads to the special double-logarithmic scaling in real space, so that the correlator shows a faster decay than any power-law when $r_x,r_y\rightarrow \infty$.

If we place the four points on the corners of a thin and long stripe by 
taking $r_x=m a$ ($a$ being the lattice spacing), this four-point correlation becomes the dipole correlation function in Eq.~\ref{dipcor}, with a power-law decay along the y-direction. However, by changing the length of the dipole, the exponent of the correlation changes rapidly, indicating that the scaling dimension of the dipole operator depends on the UV cut-off. This phenomenon, known as UV-IR mixing, is a new feature for critical liquids: the low energy behavior is sensitive to the UV `details'. In particular, the special double logarithmic scaling arises from the subsystem symmetries where the low energy spectrum contains a set of zero-modes at $k_x,k_y$ axes\cite{xu2007bond,seiberg2020exotic,paramekanti2002ring,you2021fractonic,you2019emergent}. This excessive number of low energy states at high momentum generates a set of field configurations with strong local fluctuations. As a result, we cannot simply integrate out (coarse grain) the local fluctuations nor change the UV cut-off, as the high momentum modes would bring additional singularities and hence qualitatively change the universal behavior. 

\subsection{UV-IR mixing from the perspective of global topological sectors}\label{sec:top}

Another way to visualize the UV-IR mixing is to scrutinize the \textit{topological sector} of the close-packed configurations. Our theory in Eq.\ref{fix} contains zero energy branch cuts on the $k_x,k_y$ axes that imply a subextensive (scaling as $L_x+L_y-1$) number of patterns with the same entropy. These patterns inhabit distinct topological sectors, which are connected via large gauge transformations\cite{shirley2018fractional,xu2008resonating}.

Writing the height representation of the electric field, 
\begin{align} 
E_{xy}=-\nabla_z h+\bar{E}_{xy},E_{z}=\nabla_x\nabla_y h+\bar{E}_{z},\label{dec}
\end{align}
where $\bar{E}_{z}, \bar{E}_{xy}$ are the background fields encoding the topological sector,
while the height field $h$ encodes the local fluctuations of the plaquette-dimer pattern within a sector.

\begin{figure}[h]
  \centering
      \includegraphics[width=0.4\textwidth]{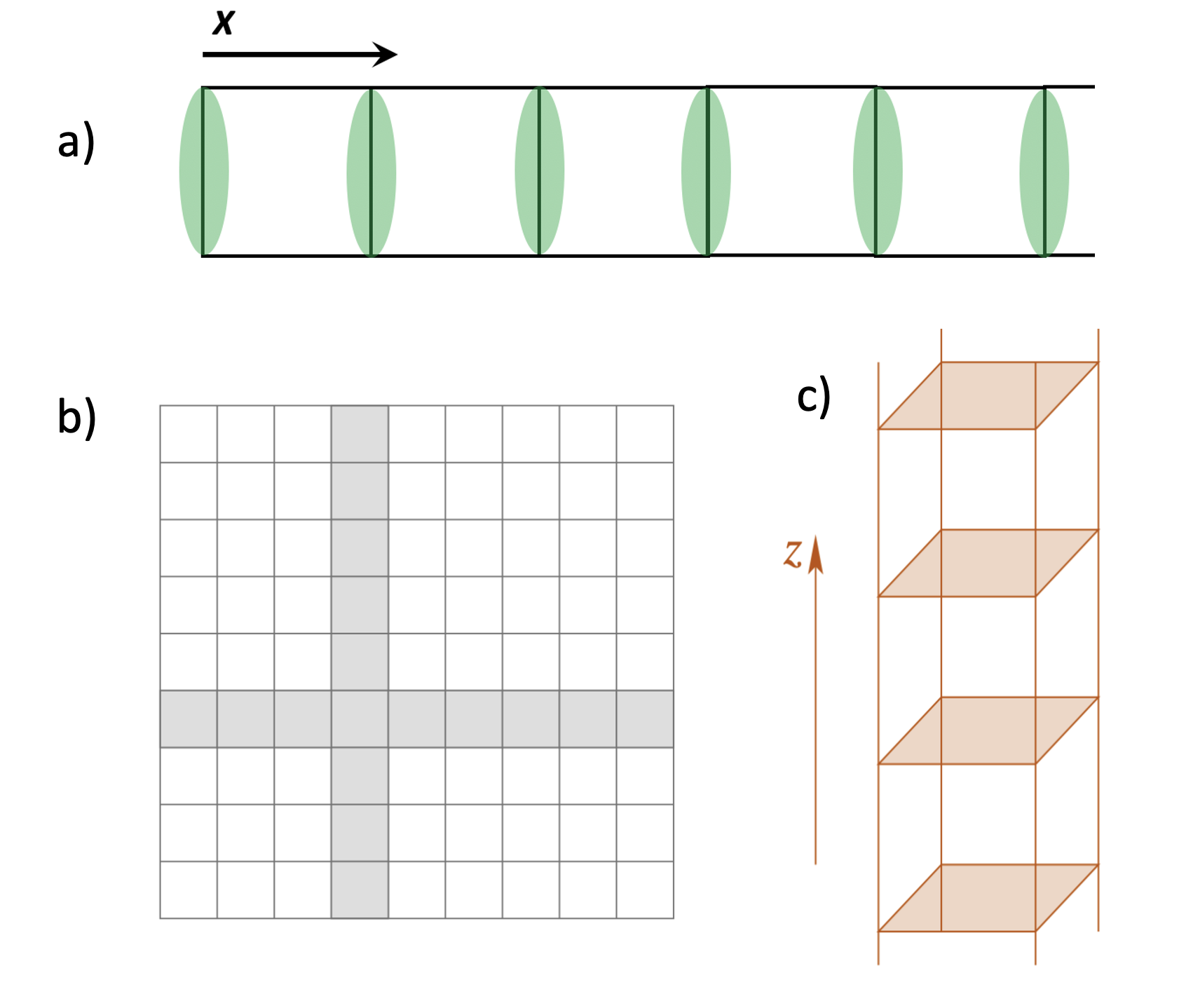}
        \caption{a) The topological sector $m_z(z,y)=\int dx ~E_z$ counts the winding number of dimers along a specific x-row. b)The view of the 3D lattice from the top onto an x-y plane. If the quantity $m_{xy}$ is fixed on all the shaded squares on a row and column,  all $m_{xy}$ are determined on the whole lattice. c) The topological sector $m_{xy}(x,y)=\int dz ~E_{xy}$ counts the winding number of plaquettes along a specific z-row.} 
\label{fig4}
\end{figure}

We can define a topological winding number on a closed manifold with periodic boundary conditions for $h$,
\begin{align} 
&m_z(z,x)=\int ~dy ~E_z ,~m_z(z,y)=\int ~dx ~E_z,\nonumber\\
&~m_{xy}(x,y)=\int dz ~E_{xy}
\end{align}
Based on the sign structure defined in Eq.~\ref{cons}, 
$m_z$ encodes the number of dimers (with a sign modulation) on the x-row(or y-row) while $m_{xy}$ encodes the plaquette number along each z-row.
These `winding numbers' characterize the topological sectors, which cannot be changed by flipping any configurations locally (Fig.~\ref{f3}). From the decomposition in Eq.~\ref{dec}, the height field representing local fluctuations does not affect the topological sector so the winding number is fixed by $\bar{E}_{z}, \bar{E}_{xy}$. 

From the Gauss-law constraint in Eq.~\ref{cons}, it follows that,
\begin{align} 
&\nabla_z m_z(z,x)=0 ,~ \nabla_z m_z(z,y)=0,\nonumber\\
&
~\nabla_x \nabla_y m_{xy}(x,y)=0
\end{align}
This identity implies that the winding numbers of the electric field along each row are not independent. In particular, if we fix the value of $m_{xy}(x,y)$ on a row and a column on the x-y plane, all other topological sectors are fixed as Fig.~\ref{fig4}. Likewise, the choices of $m_z(z,x),m_z(z,y)$ have to be independent of coordinate $z$. In addition, the choice $m_z(x),m_z(y)$ is restricted by
\begin{align} \label{wt}
\sum_x m_z(x)=\sum_y m_z(y)=W=
\int ~dy dx ~E_z .
\end{align}
Hence, there are $L_x+L_y-1$ independent values of $m_z$, each representing an independent $E_z$ winding number along the x or y column. Thus, there exists subextensive number of independent topological sectors that grows linearly with the system size. These independent sectors can be traced back to the holonomies of higher-rank gauge theories\cite{shirley2018fractional} which result in the size-dependent degeneracy of fracton topological order\cite{Haah2011-ny,vijay2017generalization,shirley2018fractional}.  

While the winding number we discuss here is reminiscent of the winding number of dimers in the 2D close-packed classical dimer liquid, the underlying physics is qualitatively different. For 2D dimers, there only exist two independent numbers, one along each orthogonal direction. This implies if the dimer's winding number along a specific row is fixed, all the other rows should display the same winding number. However, for our plaquette-dimer liquid, the numbers $m_{xy},m_{z}$ can vary on the x-y plane so the winding numbers of $m_{xy}(x_0,y_0)$ and $m_{xy}(x_0+1,y_0+1)$ are independent. 

This underpins the expression of $\bar{E}_{z}, \bar{E}_{xy}$ in terms of the winding numbers, and attribute all remaining local fluctuations to the height field.
\begin{widetext}
\begin{align} 
&\bar{E}_{xy}(x_i,y_i)=\frac{m_{xy}(x_i,y_i)}{L_z},~~\bar{E}_{z}(x_i,y_i)=\sum_x \frac{m_z(x)\delta(x-x_i)}{L_y}+\sum_y \frac{m_z(y)\delta(y-y_i)}{L_x}-\frac{W}{L_x L_y}
\label{top}
\end{align}
\end{widetext}
Here $m_z(x),m_z(y), m_{xy}(x,y)$ are the topological winding numbers along a row that are all independent of the coordinate $z$. $W$ is the total dimer winding number on the x-y plane defined as Eq.~\ref{wt}. Including summing over the different topological sectors, we can write the partition function as,
\begin{widetext}
\begin{align} 
&\mathcal{Z}=\int \mathcal{D}E_z~ \mathcal{D}E_{xy}~ e^{-\beta(E_z^2+E_{xy}^2)}=\int \mathcal{D}h~ e^{-\beta[(\partial_z h-\bar{E}_{xy})^2+(\partial_x\partial_y h- \bar{E}_{z})^2]+\alpha \cos(2\pi h)+...}
\end{align}
\end{widetext}
The fluctuations of the close-packed plaquette-dimer system is thus divided into two parts: the local fluctuation and the distinct topological sectors that require a non-local deformation along with non-contractible loops. 
Physically, the maximal value of $m_z, m_{xy}$ correspond to the staggered patterns with minimal entropy as all dimers/plaquettes are non-flippable. Likewise, $m_z, m_{xy}=0$ denotes the columnar patterns with a maximal amount of local flippability. Since the choice of $m_z, m_{xy}$ can vary in space with $L_x+L_y-1$ independent choices, we can expect configurations where the plaquettes have staggered configurations on a specific x-z slab with maximal $m_{xy}$ while at the next nearest x-z slabs, the plaquettes are in the columnar state with $m_{xy}=0$. In other words, a staggered plaquette pattern on the $y=1$ slab versus a staggered plaquette on the $y=N_y$ slab(with all other slabs exhibiting the columnar patterns) share the same flippablility and entropy. However, these two states lie in distinct topological sectors, i.e.\ cannot be connected via local fluctuations. 

This is the crucial distinction to the 
dimer liquid, whose winding is unique along each non-contractible loop: our plaquette-dimer liquid allows 'rough configurations' where the winding number can strongly fluctuate in space so the short-wavelength physics plays an important role.
For each $m_z, m_{xy}$, if we change its global flux by one unit, the entropy of the new pattern would not change up to some $1/L$ correction. Thus, the number of independent $m_z, m_{xy}$ determines the density of state engendered by the global pattern change. The branch cut at the $k_x,k_y$ axes in the Green function of the height field in Eq.~\ref{green} corresponds to the fact that there exist $L_x+L_y-1$ topological sectors whose Hilbert space cannot be connected by local fluctuations. As the `energy' in the partition function reflects the entropy of local flippable(non-flippable) patterns, if we have a zero energy line alone the momentum axes, that implies there being subextensive number of minimal energy(maximal entropy) state.  While the energy in the partition function is engendered by the entropy density due to local flips, the subextensive number of minimal energy configurations is a consequence of entropy density due to the globally distinct topological sectors.

\subsubsection{UV-IR mixing from vison conservation}\label{sec:vis}

In addition to the height field, which is desigined to incorporate the close-packeding constraint, we next provide an alternative way to comprehend the subsystem symmetry in Eq.~\ref{fix}. The higher-order derivative $(\nabla_x \nabla_y h)^2$ implies that creating a rough configuration by shifting the value of $h$ within a specific x-z(y-z) plane does not change the system entropy. 

\begin{figure}[h]
  \centering
      \includegraphics[width=0.27\textwidth]{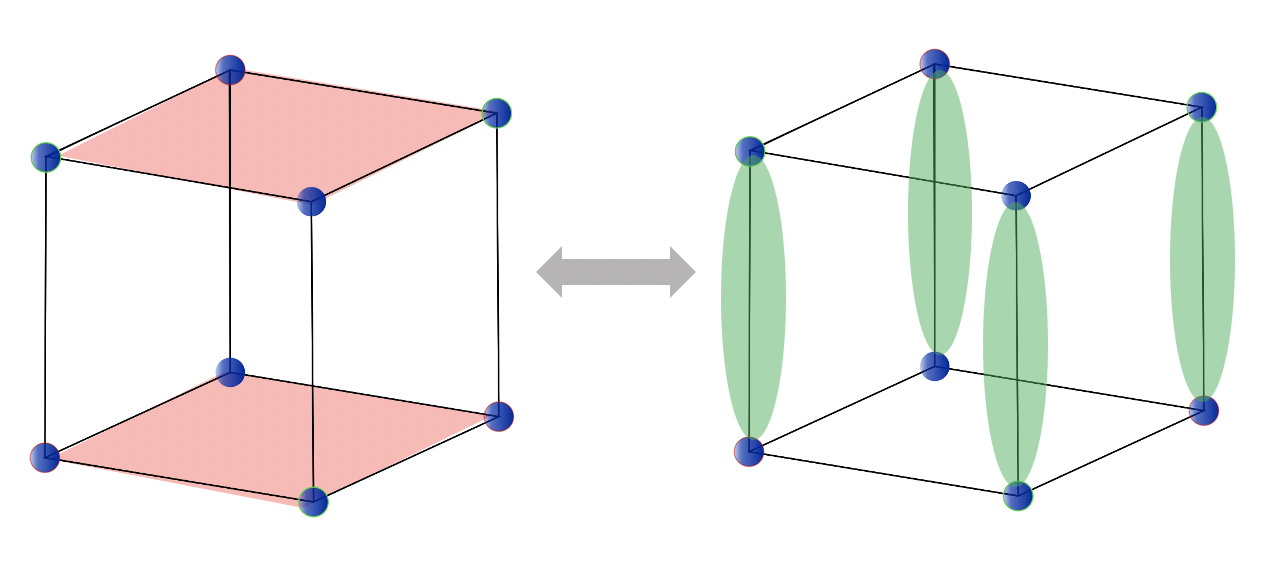}
  \caption{The vison operator $B$ living at the center of each cube flips four dimers into two plaquettes.} 
\label{f3}
\end{figure}

This fact can be readily elucidated in the vison picture. In this work, we are mainly focus on the classical version
of plaquette-dimer liquid. However, we here introduce quantum fluctuations so as to define the vison operator, but otherwise ignore quantum effects in this section.
To consider quantum fluctuations of the higher-rank electric field in Eq.\ref{cons}, we introduce the conjugate partner of the $E_{xy},E_z$, denoted as gauge potential $A_{xy},A_z$,
\begin{align} 
&A_z \rightarrow A_z- \nabla_z f, A_{xy} \rightarrow A_{xy}+\nabla_x \nabla_y f
\end{align}
The gauge potential allows a gauge transformation with $f$ being any smooth function.
The vison flux can be defined at the center of each cube as,
\begin{align} 
&B=\nabla_x \nabla_y A_z+\nabla_z A_{xy}
\end{align}
The vison operator generates a quantum resonance between distinct close packed configurations: it flips four z-dimers on the cube into two xy-plaquettes and vice versa as Fig.~\ref{f3}. Intriguingly, the vison numbers on each x-z and y-z plane are conserved,
\begin{align} 
&\int dx dz B(r)=0,~\int dz dy B(r)=0,~
\label{vison}
\end{align}
Consequently, a single vison can only move along the z-row while a pair of visons separated by an x-link can hop along the y-direction or vice versa.

In the quantum theory, the vison flux and the height field are conjugate pairs $[B(r),h(r')]=i\delta(r-r')$ so the flux operator $e^{iB(r)}$ shifts $h(r)$ by an integer. The subsystem symmetry in Eq.~\ref{th} implies shifting the value of $h$ within a specific x-z(y-z) plane does not change the system entropy. We can create such a shift by applying $e^{iq \sum_{x,z} B(x,y,z)},e^{iq \sum_{y,z} B(x,y,z)}$. Physically, each flux operator will induce a resonance between locally distinct dimer/plaquette patterns. 
However, as demonstrated in Eq.~\ref{vison}, the vison number is conserved on each x-z(y-z) plane so $\sum_{y,z} B(x,y,z)=\sum_{x,z} B(x,y,z)=0$. Thus, such a subsystem shift of the height operator does not change the physical degrees of freedom and the partition function is invariant under the planar symmetry. Alternatively, the constrained motion of the vison also suggests that the $h$ field only contains spatial fluctuation like $(\nabla_z h)^2, (\nabla_y \nabla_x h)^2$. The term $ (\nabla_x h)^2$ is not allowed as it violates flux conservation on the y-z plane.

Returning to our discussion on the topological sector in Sec.~\ref{sec:top}, it is not hard to conclude that the vison operators $B$ commute with the winding numbers $m_z,m_{xy}$. Meanwhile, there exist a set of global vison flux operators that connect between distinct topological sectors,
\begin{align} 
\int ~dx A_{xy} , \int ~dy A_{xy}, \int dz~d A_{z}
\end{align}
In the quantum higher-rank gauge theory \cite{pretko2017generalized,vijay2016fracton,shirley2018fractional,you2019fractonic}, these are the Wilson-stripe operators that uniquely define the holonomies.

\subsection{Fractonic Berezinskii-Kosterlitz-Thouless transition beyond renormalization}\label{sec:dual}

\begin{figure}[h]
  \centering
      \includegraphics[width=0.27\textwidth]{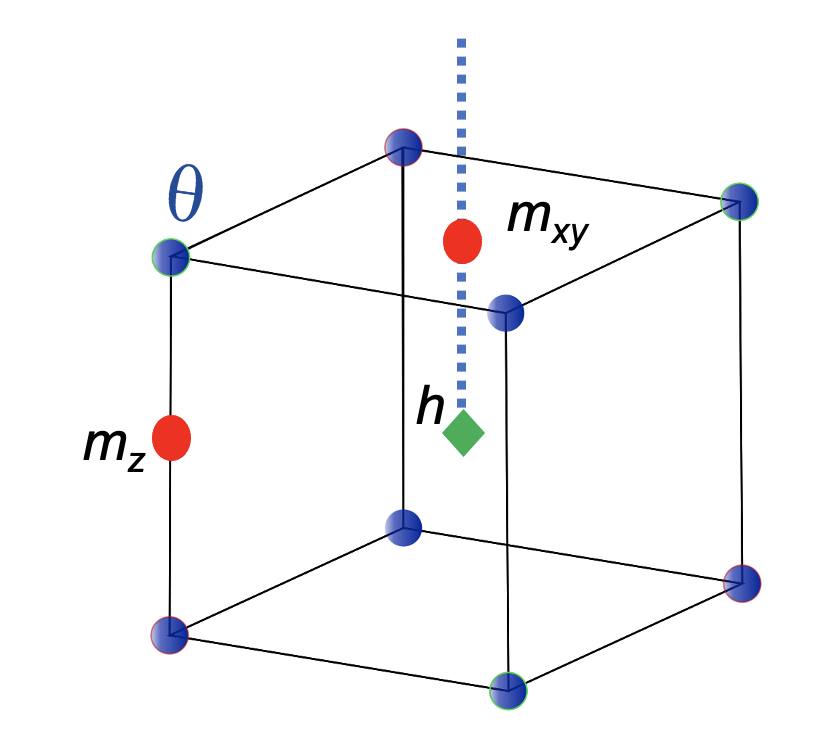}
           \caption{The rotor field $\theta$ is defined on the sites of the cube lattice, while the dual height field $h$ lives on the dual lattice at the center of each cube. Red dots:
       $m_{xy} (m_z)$ live at the center of an x-y plaquette (at the midpoint of a z-link).}
\label{fig5}
\end{figure}

In this section, we investigate the phase transition between the plaquette-dimer liquid and an ordered phase from a dual perspective. To set the stage, we consider a classical U(1) rotor model in 3D. The classical compact rotor field $\theta$ lives on the site of the cubic lattice in Fig.~\ref{fig5}.

We consider a special case where the U(1) rotor contains a ring-exchange interaction between the four rotors at the corners of each plaquette on the x-y plane in addition to the anti-ferromagnetic interaction between the two rotors along the z-link.
\begin{align} 
&\mathcal{Z}=\int \mathcal{D}\theta~ e^{-\beta(\cos(\partial_z \theta)+\cos(\partial_x\partial_y \theta))+...} \nonumber\\
&=\int \mathcal{D}\theta~ \sum_{m_z,m_p}e^{-\beta((\partial_z \theta-2\pi m_z)^2+(\partial_x\partial_y \theta-2\pi m_p)^2)+...}\label{rotor}
\end{align}
 The resulting theory contains a \textit{subsystem U(1) symmetry} $\theta \rightarrow \theta+g(x,z)+f(y,z)$ so that charges on all x-z and y-z planes are conserved.
 The compactness of the rotor field $\theta$ can be replaced by introducing two integer-valued fields $m_z,m_p$ that live on each z-link and on each x-y plaquette as Fig.~\ref{fig5}. In terms of the Villain formulation, we remove the quadratic term of $\theta$ via a Hubbard-Stratonovich (HS) transformation\cite{gorantla2021modified},
\begin{align} 
&\mathcal{Z}=\int \mathcal{D}\theta~ \mathcal{D}k_p \mathcal{D}k_z \nonumber\\
&\sum_{m_z,m_p}e^{-ik_z (\partial_z \theta-2\pi m_z)-ik_p(\partial_x\partial_y \theta-2\pi m_p)+\frac{1}{4\beta}(k_p^2+k_z^2)...}
\end{align}
Integrating out the smooth part of the $\theta$ field, the $k_z, k_p$ fields are subject to the  constraint
\begin{align} 
&\partial_z k_z-\partial_x\partial_y k_p=0~ \rightarrow ~k_p=\partial_z h,~k_z=\partial_x\partial_y h.
\end{align}
Such a constraint is identical to the close-packed constraint we derived in Eq.~\ref{cons}, so that we can rewrite the integer fields $k_z, k_p$ in terms of the height representation. 
\begin{align} 
&\mathcal{Z}=\int \mathcal{D}h~ \sum_{n}e^{-i2\pi n h+\frac{1}{4\beta}((\partial_x\partial_y h)^2+(\partial_z h)^2)...} \nonumber\\
&n=\partial_x\partial_y \partial_z \theta-\partial_z \partial_x\partial_y \theta \nonumber\\
&\rightarrow \int \mathcal{D}h~e^{\alpha \cos(2\pi h)+\frac{1}{4\beta}((\partial_x\partial_y h)^2+(\partial_z h)^2)...} 
\end{align}
This is identical to the height representation of the close-packed models we explored in Eq.~\ref{fix} with an `inverted stiffness $\frac{1}{4\beta}$'. The compactification of $\theta$ restricts the dual height field $h$ to be discrete and only takes integer values.
$n$ is the singular point defect of the $\theta$ field with $\partial_x\partial_y \partial_z \theta-\partial_z \partial_x\partial_y \theta \neq 0$ whose proliferation can bring about a disordered 'high-temperature phase' of the rotor\footnote{Such defect illustrates a configuration where a vortex line oriented along y(x) terminated at a point.}.
However, the height field has an additional subsystem symmetry $h \rightarrow h+g(x,z)+f(y,z)$, so that the total defect number $\sum n$ on each xz and yz plane should be zero. Thus, the defect anti-defect pair can only be separated along the z-direction, or one can create a four-defect quadrupole at the corners of a rectangle on the x-y plane.

Following the spirit of Villain duality\cite{kosterlitz1973ordering} in the 2D XY model, we can calculate the effective interaction between the defects by integrating out the height field at the Gaussian level,
\begin{align} \label{defectp}
&\mathcal{Z}=\sum_{n} e^{\beta V(r)n(0) n(r)},\nonumber\\
&V(0,0,z)= [\ln(z)]^2, V(x,y,0)=\infty
\end{align}
The subsystem symmetry enforces that the defects must appear in pairs along the z-direction or as quadrupoles on the x-y plane. 
Thus, the interaction strength between two defect-pairs separated in the x-y plane diverges, and the interaction strength between two defect pairs along z-direction displays `double logarithmic' scaling. 
When the defects proliferate, the entropy of the defects grows logarithmically $S\sim \ln(z)$ in the thermodynamic limit. Meanwhile, the energy cost for separating a defect pair is double logarithmic $E\sim \beta [\ln(z)]^2$ so the energy always dominates the entropy.  Consequently, the defects do not proliferate at any finite $\beta$.

Nevertheless, a higher-order defect, e.g., a dipole defect consisting of defect and anti-defect pair, could proliferate. Here we define two types of dipole densities,
\begin{align} 
&p^x(r)=x n(r)= \partial_y \partial_z \theta(r)-\partial_z \partial_y \theta(r),\nonumber\\
& p^y(r)=y n(r)= \partial_x \partial_z \theta(r)-\partial_z \partial_x \theta(r),
\end{align}
The dipole defect $p^x(p^y)$ creates a 2D vortex winding of the rotor field $\theta$ in a specific yz(xz) plane. Notably, as the rotor field exhibits a subsystem U(1) symmetry on all y-z, x-z planes, the vortices on each 2D plane are independent. In contrast to the 3D classical rotor model with a global symmetry, whose vortices are lines defects forming closed loops, the vortex defect in our model as Eq.~\ref{rotor} with subsystem symmetry can be discontinuous in space since $p^x (p^y)$ are conserved on all y-z(x-z) planes.

The interaction between the dipole defects needs to be calculated only in the planes transverse to their orientation:
\begin{align} \label{dipolep}
&V_x(r) p^x(0) p^x(r), ~V_x(0,y,z) \sim \beta/2 \ln(y^2+z^2)\nonumber\\
&V_y(r)p^y(0)  p^y(r), ~V_x(x,0,z) \sim \beta/2 \ln(x^2+z^2)
\end{align}
If we separate two dipole defect pairs $p^x$ in the same y-z plane, the energy cost grows logarithmically with distance. When these vortex dipoles proliferate, the entropy of the vortex due to its possible locations in the plane also scales as $\ln(y^2+z^2)$. 
When we tune the dipole's interaction strength by decreasing the fugacity $\beta$, the competition between energy gain and entropy production leads to a Berezinskii–Kosterlitz–Thouless (BKT) like transition that occurs at critical $\beta_c$ accompanied by the proliferation of dipole defects. This is exactly the critical point we demonstrated in Sec.~\ref{sec:one} where $\cos(2\pi \nabla_x h)$ becomes relevant.
At low temperature, $\beta > \beta_c$, the rotor $\theta$ is in the liquid phase with quasi-long range order. Its correlators are
\begin{align} 
&\langle e^{-\theta(0)\theta(z)} \rangle =e^{-\frac{1}{4\pi \beta} (\ln(z))^2}\nonumber\\
&\langle e^{-( \theta(0) \theta(e_x) \theta(0,y,z)\theta(e_x,y,z))} \rangle\nonumber\\
&=\langle e^{-(\nabla_x \theta(0)  \nabla_x \theta(0,y,z))} \rangle
=\frac{1}{(z^2+y^2)^{\frac{1}{4\pi \beta}}},\nonumber\\
&\langle e^{-( \theta(0) \theta(e_y) \theta(x,0,z)\theta(x,e_y,z))} \rangle\nonumber\\
&=\langle e^{-(\nabla_y \theta(0)  \nabla_y \theta(x,0,z))} \rangle
=\frac{1}{(z^2+x^2)^{\frac{1}{4\pi \beta}}}
\label{dipcor2}
\end{align}
Due to subsystem symmetry, the two-point function of the rotor fluid is short-range correlated while the four-point function on a thin stripe has power-law decay. In the dual language, the dual height field's correlation function has a similar form with an inverse temperature `$\frac{1}{4 \beta}$' so that the plaquette-dimer model is in the liquid phase.
In particular, in the liquid phase, the system contains both subsystem U(1) symmetry for the rotor and an \textit{emergent subsystem U(1) symmetry} for the height field due to the irrelevance of the defect operator \footnote{In the quantum version of this liquid theory\cite{gorantla2021modified}, the emergent $U(1)\times U(1)$ subsystem symmetry is responsible for the mixed 't-Hooft anomaly.}. As the liquid phase is self-dual, the monomer excitation(unpaired site in the plaquette-dimer) has the same interaction potential as the defect $n(r)$ in Eq.~\ref{defectp}. Likewise, the interaction potential between monomer dipoles grows logarithmically with distance in the same way as dipole defects in Eq.~\ref{dipolep}.

 When $\beta < \beta_c$, the dipole defects of the rotor fluid proliferate and the discreteness of $h$ from the dual theory becomes important. The proliferation of $\cos(2\pi \nabla_x h)$ engenders dipole defect $p^x,p^y$ proliferation so that the rotor is in the disordered phase. In the dual picture, the relevance of $\cos(2\pi \nabla_x  h),\cos(2\pi \nabla_y h)$ enforces the discreteness of height field so that the close-packed plaquette-dimer model is spontaneously ordered.

 The 3D phase transition  discussed here is thus reminiscent of the 2D BKT transition for classical rotor models. However, it is distinct in the following senses. 1) The BKT transition is driven by vortex proliferation while our phase transition is driven by the proliferation of higher-order defects, denoted as dipoles. 2) The BKT transition can be characterized by the renormalization group paradigm as the critical point, controlled by long wave-length modes. The critical point we discuss here carries UV-IR mixing and the critical phenomenon is controlled by short-wavelength fluctuations. This is evidenced by the fact that the height operator $\cos(2\pi h)$ at the critical point is irrelevant but its higher-order derivative $\cos(2\pi \nabla_x h)$ is relevant. As a result, the critical exponent of our phase transition does not fall into any universality and is in this sense beyond the RG paradigm. 3) Due to the subsystem symmetry, the charge of the U(1) rotor and its defect are conserved on each y-z and x-z plane so that the defects display restricted dynamics on subdimensional manifolds.

   \begin{widetext}
   
  \begin{figure}[h]  
  \centering
      \includegraphics[width=0.8\textwidth]{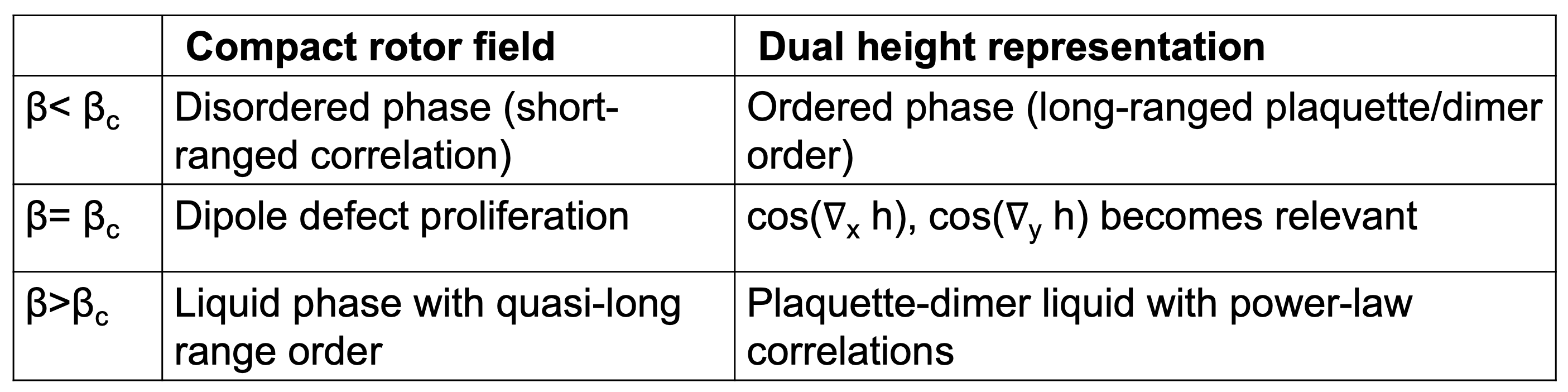}
      \caption{T-Duality table of the classical rotor model}
\label{table}
\end{figure}

 \end{widetext}
\section{Quantum fluctuations and confinement}\label{sec:qua}
We briefly touch on the quantum version of the plaquette-dimer model. To quantize the theory, we introduce the conjugate partners of the $E_{xy},E_z$, denoted as gauge potentials $A_{xy},A_z$ with the vison flux defined at the center of each cube,
\begin{align} 
&B=\nabla_x \nabla_y A_z+\nabla_z A_{xy}
\end{align}
The vison operator generates a quantum resonance between distinct close packed configurations, flipping four z-dimers on a cube into two xy-plaquettes and vice versa.  In the quantized version of the theory, the vison and the height field are canoncically conjugate $[B(r),h(r')]=i\delta(r-r')$ so that the instanton operator $e^{i2\pi h}$ creates a $2\pi$ shift of the vison. As the emergent gauge field B is compact, such an instanton event, once proliferated, can potentially lead to a confined phase. The low energy effective theory of the height field reads
\begin{align} 
&\mathcal{L}_{h}=(\partial_t h)^2+k_1 (\partial_z h)^2+k_2 (\partial_x \partial_y h)^2 .
\end{align}
The quantum theory of $h$ is defined in $3+1$D space-time. The higher-order instanton operator always has long-range correlations regardless of $k_1,k_2$,
\begin{align} 
&e^{-(\partial_x h(0)\partial_x h(r))} \rightarrow \text{Const} ~\delta(x)_{y,z \rightarrow \infty} \nonumber\\
&e^{-(\partial_y h(0)\partial_y h(r))} \rightarrow \text{Const} ~\delta(y)_{x,z \rightarrow \infty}
\end{align}
The delta function here again arises from the subsystem symmetry and the resulting vison number conserved in the $x-z,y-z$ planes. As the instantons are always relevant, the system does not yield a deconfined phase and a quantum dimer-plaquette liquid phase does not exist in this framework.

\section{Summary and outlook}

We hope that our work has established a new class of mixed hardcore plaquette-dimer models as an interesting object of study. The perhaps simplest model in this class, for the anisotropic cubic lattice, the main subject of this work, has already turned out to harbour rich promise of `fractonic' and post-RG critical phenomenology. 

Our analysis has been primarily field-theoretic in nature. As a next step, it is clearly desirable to undertake a detailed numerical study of the microscopic model system. This could determine the correlations present in the ensemble of hardcore plaquette-dimer coverings, in particular, which phase it actually realises. The addition of appropriate interactions could then drive it across the phase transition we have analysed, in analogy to the study  in Ref.~\onlinecite{2005PhRvL..94w5702A} on the square lattice dimer model. 

More braodly, it is clearly desirable to undertake a systematic study of variants of the model analysed here. We close this account by venturing the first few steps in this direction.

\subsection{Close-packed trimer-dimer state: Classical type-II fractons}

This takes us to a discussion of a possible type-II fracton liquid in an analogous close-packed trimer-dimer system. We will not study this model in detail but introduce the basic setup and leave a detailed study for future exploration.

\begin{figure}[h]
  \centering
      \includegraphics[width=0.5\textwidth]{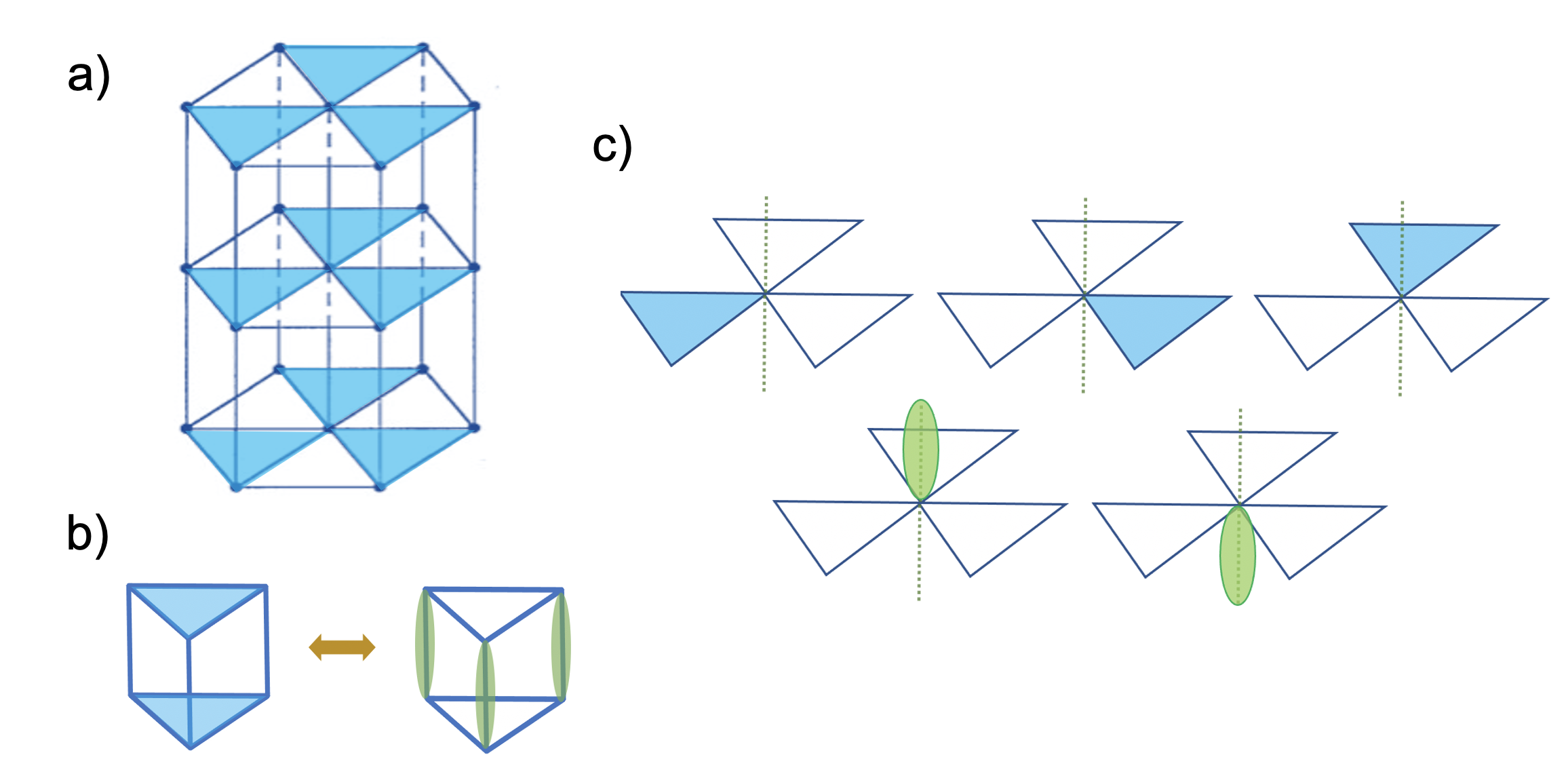}
      \caption{a) The layered triangular lattice. b) A local resonance between two distinct patterns. c) Five possible trimer-dimer closed-pack patterns adjacent to a site.}
\label{tri1}
\end{figure}

We consider a layered triangular lattice with each site part of two bonds along the z-direction and three downward triangles in the x-y plane. The system only allows  configurations that have each site either connected to a trimer on one of the three downward triangles, or to a dimer on one of the two z-bonds, as illustrated as Fig.~\ref{tri1}. Such a constrained space of close-packed dimer-trimer configurations has extensive degeneracy (non-vanishing entropy). Fig.~\ref{tri1}-b illustrates a  fluctuation that resonates two trimers into three dimers within a unit prism.

  To analyse the properties of the ensemble of these close-packed patterns, we resolve the local dimer-trimer constraint by representing trimer and dimer coverage via a higher-rank electric field\cite{bulmash2018generalized,gromov2019towards},
  \begin{align} 
E_{t}=\eta_t T_{xy}, ~E_{z}=\eta_z D_{z}. \label{def}
\end{align}
$E_{t}$ is defined on the center of the downward triangles of the x-y plane while $E_z$ lives on each z-link.
$T_{xy},  D_{z}$ refer to the number of trimers and dimers  in the x-y plane and on z-link, respectively. The $\eta_t, \eta_z$ are the lattice staggering factors with a sign structure defined in Fig.~\ref{tri1}. Based on this notation, the dimer-trimer constraint can be interpreted as a Gauss-law 
\begin{align} 
(a\Delta_x\Delta_y+\Delta_y+1/a) E_{t}+\Delta_z E_{z}=\eta_q (1-Q)\ , \label{type2}
\end{align}
where $Q$ denotes the monomer number on a site. The close-packed configurations have $Q=0$ and the staggering background charge $\eta_q$ (illustrated in Fig.~\ref{tri1}), thus obeying the hardcore constraint of each site being either connected to a dimer or a trimer. 
The close-packed dimer-trimer constraint in Eq.~\ref{type2} is reminiscent of the quantum Newmann-Moore\cite{newman_1,newman_2,zhou2021fractal} model and type-II fracton gauge theory in Ref.\cite{bulmash2018generalized,gromov2019towards} whose the charge exhibits additional conservation law on a 2D fractal.

  \begin{figure}[h]
  \centering
      \includegraphics[width=0.45\textwidth]{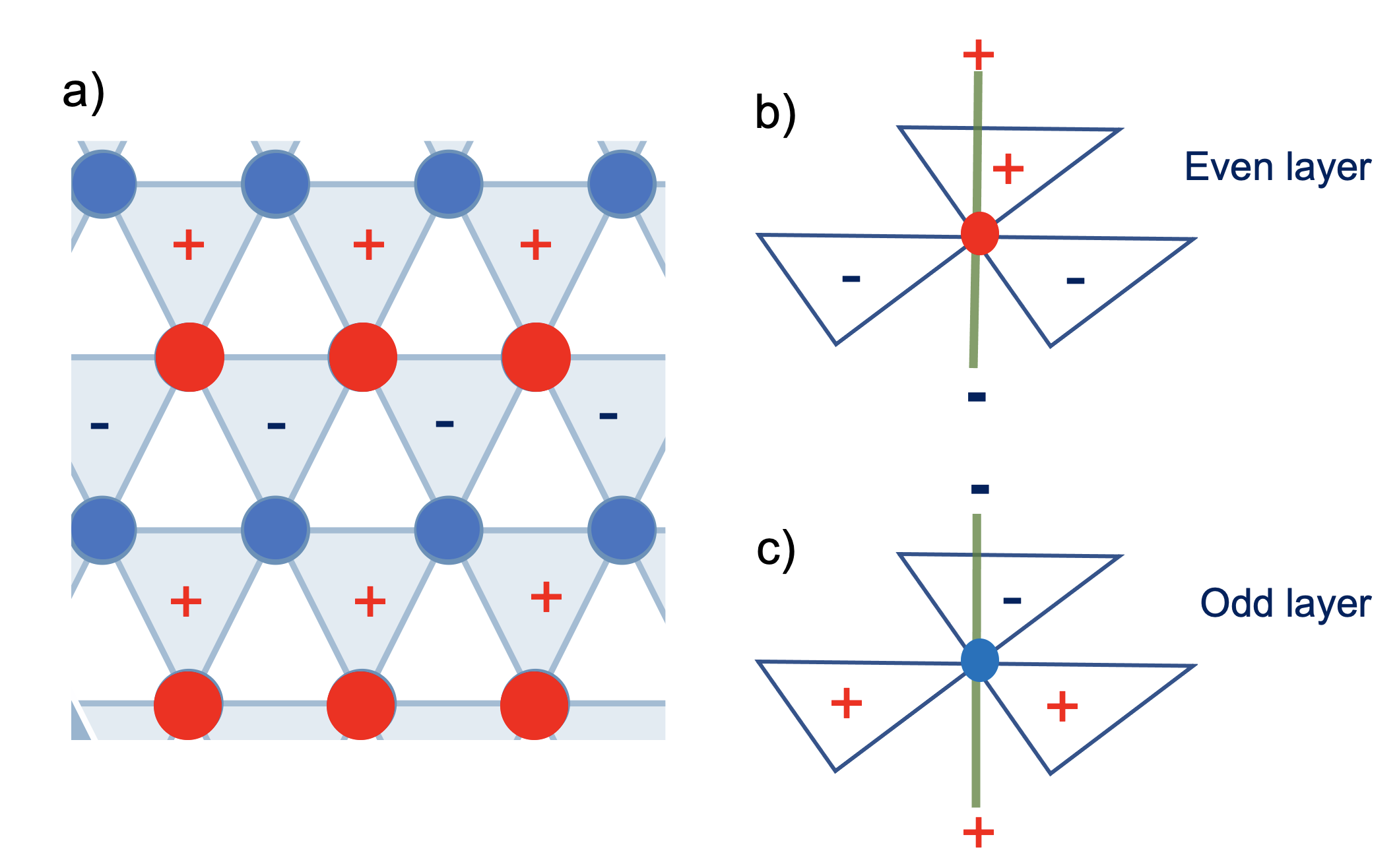}
      \caption{a) Sign factors $\eta_t ,\eta_q$ on even layers. Red denotes positive while blue denotes negative. Odd layers have exactly opposite sign factors. b) Illustration of the sign factor $\eta_t ,\eta_q,\eta_z$ near each site.}
\label{tri2}
\end{figure}

To satisfy this local constraint, one can parameterize the electric field as
\begin{align} 
E_{t}=-\Delta_z h+\bar{E}_{t},E_{z}=(a\Delta_x\Delta_y+\Delta_y+1/a) h+\bar{E}_{z},
\end{align}
Here $h$ lives at the center of each prism and $\bar{E}_{z},\bar{E}_{t}$ are the background patterns. If we consider the close-packing problem as a hardcore constraint, the partition function contains a summation over all the trimer/dimer configurations with equal Boltzmann weights. We anticipate the possibility of a similar high-entropy trimer-dimer liquid phase. In particular, UV-IR mixing is again expected as the low energy fluctuations in this theory are controlled by short wave-length physics and the liquid-to-ordered transition should again lie outside the renormalization group scheme. 

\emph{Acknowledgement}--- 
We are grateful for many useful exchanges with S.L. Sondhi. YY acknowledges the support of the Max Planck Institute for the Physics of Complex Systems, where this work was initated.
We acknowledge support from the Deutsche 
Forschungsgemeinschaft through SFB 1143 (project-id 247310070) and cluster of excellence 
ct.qmat (EXC 2147, project-id 390858490).


\begin{thebibliography}{58}%
\makeatletter
\providecommand \@ifxundefined [1]{%
 \@ifx{#1\undefined}
}%
\providecommand \@ifnum [1]{%
 \ifnum #1\expandafter \@firstoftwo
 \else \expandafter \@secondoftwo
 \fi
}%
\providecommand \@ifx [1]{%
 \ifx #1\expandafter \@firstoftwo
 \else \expandafter \@secondoftwo
 \fi
}%
\providecommand \natexlab [1]{#1}%
\providecommand \enquote  [1]{``#1''}%
\providecommand \bibnamefont  [1]{#1}%
\providecommand \bibfnamefont [1]{#1}%
\providecommand \citenamefont [1]{#1}%
\providecommand \href@noop [0]{\@secondoftwo}%
\providecommand \href [0]{\begingroup \@sanitize@url \@href}%
\providecommand \@href[1]{\@@startlink{#1}\@@href}%
\providecommand \@@href[1]{\endgroup#1\@@endlink}%
\providecommand \@sanitize@url [0]{\catcode `\\12\catcode `\$12\catcode
  `\&12\catcode `\#12\catcode `\^12\catcode `\_12\catcode `\%12\relax}%
\providecommand \@@startlink[1]{}%
\providecommand \@@endlink[0]{}%
\providecommand \url  [0]{\begingroup\@sanitize@url \@url }%
\providecommand \@url [1]{\endgroup\@href {#1}{\urlprefix }}%
\providecommand \urlprefix  [0]{URL }%
\providecommand \Eprint [0]{\href }%
\providecommand \doibase [0]{https://doi.org/}%
\providecommand \selectlanguage [0]{\@gobble}%
\providecommand \bibinfo  [0]{\@secondoftwo}%
\providecommand \bibfield  [0]{\@secondoftwo}%
\providecommand \translation [1]{[#1]}%
\providecommand \BibitemOpen [0]{}%
\providecommand \bibitemStop [0]{}%
\providecommand \bibitemNoStop [0]{.\EOS\space}%
\providecommand \EOS [0]{\spacefactor3000\relax}%
\providecommand \BibitemShut  [1]{\csname bibitem#1\endcsname}%
\let\auto@bib@innerbib\@empty
\bibitem [{\citenamefont {Kasteleyn}(1961)}]{kasteleyn1961statistics}%
  \BibitemOpen
  \bibfield  {author} {\bibinfo {author} {\bibfnamefont {P.~W.}\ \bibnamefont
  {Kasteleyn}},\ }\bibfield  {title} {\bibinfo {title} {The statistics of
  dimers on a lattice: I. the number of dimer arrangements on a quadratic
  lattice},\ }\href@noop {} {\bibfield  {journal} {\bibinfo  {journal}
  {Physica}\ }\textbf {\bibinfo {volume} {27}},\ \bibinfo {pages} {1209}
  (\bibinfo {year} {1961})}\BibitemShut {NoStop}%
\bibitem [{\citenamefont {Kasteleyn}(1963)}]{kasteleyn1963dimer}%
  \BibitemOpen
  \bibfield  {author} {\bibinfo {author} {\bibfnamefont {P.~W.}\ \bibnamefont
  {Kasteleyn}},\ }\bibfield  {title} {\bibinfo {title} {Dimer statistics and
  phase transitions},\ }\href@noop {} {\bibfield  {journal} {\bibinfo
  {journal} {Journal of Mathematical Physics}\ }\textbf {\bibinfo {volume}
  {4}},\ \bibinfo {pages} {287} (\bibinfo {year} {1963})}\BibitemShut {NoStop}%
\bibitem [{\citenamefont {Fisher}\ and\ \citenamefont
  {Stephenson}(1963)}]{fisher1963statistical}%
  \BibitemOpen
  \bibfield  {author} {\bibinfo {author} {\bibfnamefont {M.~E.}\ \bibnamefont
  {Fisher}}\ and\ \bibinfo {author} {\bibfnamefont {J.}~\bibnamefont
  {Stephenson}},\ }\bibfield  {title} {\bibinfo {title} {Statistical mechanics
  of dimers on a plane lattice. ii. dimer correlations and monomers},\
  }\href@noop {} {\bibfield  {journal} {\bibinfo  {journal} {Physical Review}\
  }\textbf {\bibinfo {volume} {132}},\ \bibinfo {pages} {1411} (\bibinfo {year}
  {1963})}\BibitemShut {NoStop}%
\bibitem [{\citenamefont {Rokhsar}\ and\ \citenamefont
  {Kivelson}(1988)}]{rokhsar1988superconductivity}%
  \BibitemOpen
  \bibfield  {author} {\bibinfo {author} {\bibfnamefont {D.~S.}\ \bibnamefont
  {Rokhsar}}\ and\ \bibinfo {author} {\bibfnamefont {S.~A.}\ \bibnamefont
  {Kivelson}},\ }\bibfield  {title} {\bibinfo {title} {Superconductivity and
  the quantum hard-core dimer gas},\ }\href@noop {} {\bibfield  {journal}
  {\bibinfo  {journal} {Physical review letters}\ }\textbf {\bibinfo {volume}
  {61}},\ \bibinfo {pages} {2376} (\bibinfo {year} {1988})}\BibitemShut
  {NoStop}%
\bibitem [{\citenamefont {Henley}(1997)}]{henley1997relaxation}%
  \BibitemOpen
  \bibfield  {author} {\bibinfo {author} {\bibfnamefont {C.~L.}\ \bibnamefont
  {Henley}},\ }\bibfield  {title} {\bibinfo {title} {Relaxation time for a
  dimer covering with height representation},\ }\href@noop {} {\bibfield
  {journal} {\bibinfo  {journal} {Journal of statistical physics}\ }\textbf
  {\bibinfo {volume} {89}},\ \bibinfo {pages} {483} (\bibinfo {year}
  {1997})}\BibitemShut {NoStop}%
\bibitem [{\citenamefont {Henley}(2004)}]{henley2004classical}%
  \BibitemOpen
  \bibfield  {author} {\bibinfo {author} {\bibfnamefont {C.~L.}\ \bibnamefont
  {Henley}},\ }\bibfield  {title} {\bibinfo {title} {From classical to quantum
  dynamics at rokhsar--kivelson points},\ }\href@noop {} {\bibfield  {journal}
  {\bibinfo  {journal} {Journal of Physics: Condensed Matter}\ }\textbf
  {\bibinfo {volume} {16}},\ \bibinfo {pages} {S891} (\bibinfo {year}
  {2004})}\BibitemShut {NoStop}%
\bibitem [{\citenamefont {Moessner}\ and\ \citenamefont
  {Raman}(2011)}]{moessner2011quantum}%
  \BibitemOpen
  \bibfield  {author} {\bibinfo {author} {\bibfnamefont {R.}~\bibnamefont
  {Moessner}}\ and\ \bibinfo {author} {\bibfnamefont {K.~S.}\ \bibnamefont
  {Raman}},\ }\bibfield  {title} {\bibinfo {title} {Quantum dimer models},\
  }in\ \href@noop {} {\emph {\bibinfo {booktitle} {Introduction to Frustrated
  Magnetism}}}\ (\bibinfo  {publisher} {Springer},\ \bibinfo {year} {2011})\
  pp.\ \bibinfo {pages} {437--479}\BibitemShut {NoStop}%
\bibitem [{\citenamefont {Moessner}\ and\ \citenamefont
  {Sondhi}(2001)}]{moessner2001resonating}%
  \BibitemOpen
  \bibfield  {author} {\bibinfo {author} {\bibfnamefont {R.}~\bibnamefont
  {Moessner}}\ and\ \bibinfo {author} {\bibfnamefont {S.~L.}\ \bibnamefont
  {Sondhi}},\ }\bibfield  {title} {\bibinfo {title} {Resonating valence bond
  phase in the triangular lattice quantum dimer model},\ }\href@noop {}
  {\bibfield  {journal} {\bibinfo  {journal} {Physical Review Letters}\
  }\textbf {\bibinfo {volume} {86}},\ \bibinfo {pages} {1881} (\bibinfo {year}
  {2001})}\BibitemShut {NoStop}%
\bibitem [{\citenamefont {Castelnovo}\ \emph {et~al.}(2008)\citenamefont
  {Castelnovo}, \citenamefont {Moessner},\ and\ \citenamefont
  {Sondhi}}]{castelnovo2008magnetic}%
  \BibitemOpen
  \bibfield  {author} {\bibinfo {author} {\bibfnamefont {C.}~\bibnamefont
  {Castelnovo}}, \bibinfo {author} {\bibfnamefont {R.}~\bibnamefont
  {Moessner}},\ and\ \bibinfo {author} {\bibfnamefont {S.~L.}\ \bibnamefont
  {Sondhi}},\ }\bibfield  {title} {\bibinfo {title} {Magnetic monopoles in spin
  ice},\ }\href@noop {} {\bibfield  {journal} {\bibinfo  {journal} {Nature}\
  }\textbf {\bibinfo {volume} {451}},\ \bibinfo {pages} {42} (\bibinfo {year}
  {2008})}\BibitemShut {NoStop}%
\bibitem [{\citenamefont {{Yokoi}}\ \emph {et~al.}(1986)\citenamefont
  {{Yokoi}}, \citenamefont {{Nagle}},\ and\ \citenamefont
  {{Salinas}}}]{1986JSP....44..729Y}%
  \BibitemOpen
  \bibfield  {author} {\bibinfo {author} {\bibfnamefont {C.~S.~O.}\
  \bibnamefont {{Yokoi}}}, \bibinfo {author} {\bibfnamefont {J.~F.}\
  \bibnamefont {{Nagle}}},\ and\ \bibinfo {author} {\bibfnamefont {S.~R.}\
  \bibnamefont {{Salinas}}},\ }\bibfield  {title} {\bibinfo {title} {{Dimer
  pair correlations on the brick lattice}},\ }\href
  {https://doi.org/10.1007/BF01011905} {\bibfield  {journal} {\bibinfo
  {journal} {Journal of Statistical Physics}\ }\textbf {\bibinfo {volume}
  {44}},\ \bibinfo {pages} {729} (\bibinfo {year} {1986})}\BibitemShut
  {NoStop}%
\bibitem [{\citenamefont {{Zeng}}\ and\ \citenamefont
  {{Henley}}(1997)}]{1997PhRvB..5514935Z}%
  \BibitemOpen
  \bibfield  {author} {\bibinfo {author} {\bibfnamefont {C.}~\bibnamefont
  {{Zeng}}}\ and\ \bibinfo {author} {\bibfnamefont {C.~L.}\ \bibnamefont
  {{Henley}}},\ }\bibfield  {title} {\bibinfo {title} {{Zero-temperature phase
  transitions of an antiferromagnetic Ising model of general spin on a
  triangular lattice}},\ }\href {https://doi.org/10.1103/PhysRevB.55.14935}
  {\bibfield  {journal} {\bibinfo  {journal} {\prb}\ }\textbf {\bibinfo
  {volume} {55}},\ \bibinfo {pages} {14935} (\bibinfo {year} {1997})},\ \Eprint
  {https://arxiv.org/abs/cond-mat/9609007} {arXiv:cond-mat/9609007 [cond-mat]}
  \BibitemShut {NoStop}%
\bibitem [{\citenamefont {{Alet}}\ \emph {et~al.}(2005)\citenamefont {{Alet}},
  \citenamefont {{Jacobsen}}, \citenamefont {{Misguich}}, \citenamefont
  {{Pasquier}}, \citenamefont {{Mila}},\ and\ \citenamefont
  {{Troyer}}}]{2005PhRvL..94w5702A}%
  \BibitemOpen
  \bibfield  {author} {\bibinfo {author} {\bibfnamefont {F.}~\bibnamefont
  {{Alet}}}, \bibinfo {author} {\bibfnamefont {J.~L.}\ \bibnamefont
  {{Jacobsen}}}, \bibinfo {author} {\bibfnamefont {G.}~\bibnamefont
  {{Misguich}}}, \bibinfo {author} {\bibfnamefont {V.}~\bibnamefont
  {{Pasquier}}}, \bibinfo {author} {\bibfnamefont {F.}~\bibnamefont {{Mila}}},\
  and\ \bibinfo {author} {\bibfnamefont {M.}~\bibnamefont {{Troyer}}},\
  }\bibfield  {title} {\bibinfo {title} {{Interacting Classical Dimers on the
  Square Lattice}},\ }\href {https://doi.org/10.1103/PhysRevLett.94.235702}
  {\bibfield  {journal} {\bibinfo  {journal} {\prl}\ }\textbf {\bibinfo
  {volume} {94}},\ \bibinfo {eid} {235702} (\bibinfo {year} {2005})},\ \Eprint
  {https://arxiv.org/abs/cond-mat/0501241} {arXiv:cond-mat/0501241
  [cond-mat.stat-mech]} \BibitemShut {NoStop}%
\bibitem [{\citenamefont {{Huse}}\ \emph {et~al.}(2003)\citenamefont {{Huse}},
  \citenamefont {{Krauth}}, \citenamefont {{Moessner}},\ and\ \citenamefont
  {{Sondhi}}}]{2003PhRvL..91p7004H}%
  \BibitemOpen
  \bibfield  {author} {\bibinfo {author} {\bibfnamefont {D.~A.}\ \bibnamefont
  {{Huse}}}, \bibinfo {author} {\bibfnamefont {W.}~\bibnamefont {{Krauth}}},
  \bibinfo {author} {\bibfnamefont {R.}~\bibnamefont {{Moessner}}},\ and\
  \bibinfo {author} {\bibfnamefont {S.~L.}\ \bibnamefont {{Sondhi}}},\
  }\bibfield  {title} {\bibinfo {title} {{Coulomb and Liquid Dimer Models in
  Three Dimensions}},\ }\href {https://doi.org/10.1103/PhysRevLett.91.167004}
  {\bibfield  {journal} {\bibinfo  {journal} {\prl}\ }\textbf {\bibinfo
  {volume} {91}},\ \bibinfo {eid} {167004} (\bibinfo {year} {2003})},\ \Eprint
  {https://arxiv.org/abs/cond-mat/0305318} {arXiv:cond-mat/0305318
  [cond-mat.stat-mech]} \BibitemShut {NoStop}%
\bibitem [{\citenamefont {{Nienhuis}}\ \emph {et~al.}(1984)\citenamefont
  {{Nienhuis}}, \citenamefont {{Hilhorst}},\ and\ \citenamefont
  {{Bl{\"o}te}}}]{1984JPhA...17.3559N}%
  \BibitemOpen
  \bibfield  {author} {\bibinfo {author} {\bibfnamefont {B.}~\bibnamefont
  {{Nienhuis}}}, \bibinfo {author} {\bibfnamefont {H.~J.}\ \bibnamefont
  {{Hilhorst}}},\ and\ \bibinfo {author} {\bibfnamefont {H.~W.~J.}\
  \bibnamefont {{Bl{\"o}te}}},\ }\bibfield  {title} {\bibinfo {title}
  {{Triangular SOS models and cubic-crystal shapes}},\ }\href
  {https://doi.org/10.1088/0305-4470/17/18/025} {\bibfield  {journal} {\bibinfo
   {journal} {Journal of Physics A Mathematical General}\ }\textbf {\bibinfo
  {volume} {17}},\ \bibinfo {pages} {3559} (\bibinfo {year}
  {1984})}\BibitemShut {NoStop}%
\bibitem [{\citenamefont {Moessner}\ \emph {et~al.}(2001)\citenamefont
  {Moessner}, \citenamefont {Sondhi},\ and\ \citenamefont
  {Fradkin}}]{moessner2001short}%
  \BibitemOpen
  \bibfield  {author} {\bibinfo {author} {\bibfnamefont {R.}~\bibnamefont
  {Moessner}}, \bibinfo {author} {\bibfnamefont {S.~L.}\ \bibnamefont
  {Sondhi}},\ and\ \bibinfo {author} {\bibfnamefont {E.}~\bibnamefont
  {Fradkin}},\ }\bibfield  {title} {\bibinfo {title} {Short-ranged resonating
  valence bond physics, quantum dimer models, and ising gauge theories},\
  }\href@noop {} {\bibfield  {journal} {\bibinfo  {journal} {Physical Review
  B}\ }\textbf {\bibinfo {volume} {65}},\ \bibinfo {pages} {024504} (\bibinfo
  {year} {2001})}\BibitemShut {NoStop}%
\bibitem [{\citenamefont {{Moessner}}\ and\ \citenamefont
  {{Sondhi}}(2003)}]{2003PhRvB..68r4512M}%
  \BibitemOpen
  \bibfield  {author} {\bibinfo {author} {\bibfnamefont {R.}~\bibnamefont
  {{Moessner}}}\ and\ \bibinfo {author} {\bibfnamefont {S.~L.}\ \bibnamefont
  {{Sondhi}}},\ }\bibfield  {title} {\bibinfo {title} {{Three-dimensional
  resonating-valence-bond liquids and their excitations}},\ }\href
  {https://doi.org/10.1103/PhysRevB.68.184512} {\bibfield  {journal} {\bibinfo
  {journal} {\prb}\ }\textbf {\bibinfo {volume} {68}},\ \bibinfo {eid} {184512}
  (\bibinfo {year} {2003})},\ \Eprint {https://arxiv.org/abs/cond-mat/0307592}
  {arXiv:cond-mat/0307592 [cond-mat.str-el]} \BibitemShut {NoStop}%
\bibitem [{\citenamefont {Hermele}\ \emph {et~al.}(2004)\citenamefont
  {Hermele}, \citenamefont {Fisher},\ and\ \citenamefont
  {Balents}}]{hermele2004pyrochlore}%
  \BibitemOpen
  \bibfield  {author} {\bibinfo {author} {\bibfnamefont {M.}~\bibnamefont
  {Hermele}}, \bibinfo {author} {\bibfnamefont {M.~P.}\ \bibnamefont
  {Fisher}},\ and\ \bibinfo {author} {\bibfnamefont {L.}~\bibnamefont
  {Balents}},\ }\bibfield  {title} {\bibinfo {title} {Pyrochlore photons: The u
  (1) spin liquid in a s= 1 2 three-dimensional frustrated magnet},\
  }\href@noop {} {\bibfield  {journal} {\bibinfo  {journal} {Physical Review
  B}\ }\textbf {\bibinfo {volume} {69}},\ \bibinfo {pages} {064404} (\bibinfo
  {year} {2004})}\BibitemShut {NoStop}%
\bibitem [{\citenamefont {Flicker}\ \emph {et~al.}(2020)\citenamefont
  {Flicker}, \citenamefont {Simon},\ and\ \citenamefont
  {Parameswaran}}]{flicker2020classical}%
  \BibitemOpen
  \bibfield  {author} {\bibinfo {author} {\bibfnamefont {F.}~\bibnamefont
  {Flicker}}, \bibinfo {author} {\bibfnamefont {S.~H.}\ \bibnamefont {Simon}},\
  and\ \bibinfo {author} {\bibfnamefont {S.}~\bibnamefont {Parameswaran}},\
  }\bibfield  {title} {\bibinfo {title} {Classical dimers on penrose tilings},\
  }\href@noop {} {\bibfield  {journal} {\bibinfo  {journal} {Physical Review
  X}\ }\textbf {\bibinfo {volume} {10}},\ \bibinfo {pages} {011005} (\bibinfo
  {year} {2020})}\BibitemShut {NoStop}%
\bibitem [{\citenamefont {{Ramola}}\ \emph {et~al.}(2015)\citenamefont
  {{Ramola}}, \citenamefont {{Damle}},\ and\ \citenamefont
  {{Dhar}}}]{2015PhRvL.114s0601R}%
  \BibitemOpen
  \bibfield  {author} {\bibinfo {author} {\bibfnamefont {K.}~\bibnamefont
  {{Ramola}}}, \bibinfo {author} {\bibfnamefont {K.}~\bibnamefont {{Damle}}},\
  and\ \bibinfo {author} {\bibfnamefont {D.}~\bibnamefont {{Dhar}}},\
  }\bibfield  {title} {\bibinfo {title} {{Columnar Order and Ashkin-Teller
  Criticality in Mixtures of Hard Squares and Dimers}},\ }\href
  {https://doi.org/10.1103/PhysRevLett.114.190601} {\bibfield  {journal}
  {\bibinfo  {journal} {\prl}\ }\textbf {\bibinfo {volume} {114}},\ \bibinfo
  {eid} {190601} (\bibinfo {year} {2015})},\ \Eprint
  {https://arxiv.org/abs/1408.4943} {arXiv:1408.4943 [cond-mat.stat-mech]}
  \BibitemShut {NoStop}%
\bibitem [{\citenamefont {Xu}\ and\ \citenamefont
  {Wu}(2008)}]{xu2008resonating}%
  \BibitemOpen
  \bibfield  {author} {\bibinfo {author} {\bibfnamefont {C.}~\bibnamefont
  {Xu}}\ and\ \bibinfo {author} {\bibfnamefont {C.}~\bibnamefont {Wu}},\
  }\bibfield  {title} {\bibinfo {title} {Resonating plaquette phases in
  {S}{U}(4) {H}eisenberg antiferromagnet},\ }\href
  {https://doi.org/10.1103/PhysRevB.77.134449} {\bibfield  {journal} {\bibinfo
  {journal} {Phys. Rev. B}\ }\textbf {\bibinfo {volume} {77}},\ \bibinfo
  {pages} {134449} (\bibinfo {year} {2008})}\BibitemShut {NoStop}%
\bibitem [{\citenamefont {Pankov}\ \emph {et~al.}(2007)\citenamefont {Pankov},
  \citenamefont {Moessner},\ and\ \citenamefont
  {Sondhi}}]{pankov2007resonating}%
  \BibitemOpen
  \bibfield  {author} {\bibinfo {author} {\bibfnamefont {S.}~\bibnamefont
  {Pankov}}, \bibinfo {author} {\bibfnamefont {R.}~\bibnamefont {Moessner}},\
  and\ \bibinfo {author} {\bibfnamefont {S.~L.}\ \bibnamefont {Sondhi}},\
  }\bibfield  {title} {\bibinfo {title} {Resonating singlet valence
  plaquettes},\ }\href@noop {} {\bibfield  {journal} {\bibinfo  {journal}
  {Physical Review B}\ }\textbf {\bibinfo {volume} {76}},\ \bibinfo {pages}
  {104436} (\bibinfo {year} {2007})}\BibitemShut {NoStop}%
\bibitem [{\citenamefont {{Vigneshwar}}\ \emph {et~al.}(2019)\citenamefont
  {{Vigneshwar}}, \citenamefont {{Mandal}}, \citenamefont {{Damle}},
  \citenamefont {{Dhar}},\ and\ \citenamefont
  {{Rajesh}}}]{2019PhRvE..99e2129V}%
  \BibitemOpen
  \bibfield  {author} {\bibinfo {author} {\bibfnamefont {N.}~\bibnamefont
  {{Vigneshwar}}}, \bibinfo {author} {\bibfnamefont {D.}~\bibnamefont
  {{Mandal}}}, \bibinfo {author} {\bibfnamefont {K.}~\bibnamefont {{Damle}}},
  \bibinfo {author} {\bibfnamefont {D.}~\bibnamefont {{Dhar}}},\ and\ \bibinfo
  {author} {\bibfnamefont {R.}~\bibnamefont {{Rajesh}}},\ }\bibfield  {title}
  {\bibinfo {title} {{Phase diagram of a system of hard cubes on the cubic
  lattice}},\ }\href {https://doi.org/10.1103/PhysRevE.99.052129} {\bibfield
  {journal} {\bibinfo  {journal} {\pre}\ }\textbf {\bibinfo {volume} {99}},\
  \bibinfo {eid} {052129} (\bibinfo {year} {2019})},\ \Eprint
  {https://arxiv.org/abs/1902.06408} {arXiv:1902.06408 [cond-mat.stat-mech]}
  \BibitemShut {NoStop}%
\bibitem [{\citenamefont {You}\ \emph {et~al.}(2020{\natexlab{a}})\citenamefont
  {You}, \citenamefont {Bi},\ and\ \citenamefont {Pretko}}]{you2019emergent}%
  \BibitemOpen
  \bibfield  {author} {\bibinfo {author} {\bibfnamefont {Y.}~\bibnamefont
  {You}}, \bibinfo {author} {\bibfnamefont {Z.}~\bibnamefont {Bi}},\ and\
  \bibinfo {author} {\bibfnamefont {M.}~\bibnamefont {Pretko}},\ }\bibfield
  {title} {\bibinfo {title} {Emergent fractons and algebraic quantum liquid
  from plaquette melting transitions},\ }\href
  {https://doi.org/10.1103/PhysRevResearch.2.013162} {\bibfield  {journal}
  {\bibinfo  {journal} {Phys. Rev. Research}\ }\textbf {\bibinfo {volume}
  {2}},\ \bibinfo {pages} {013162} (\bibinfo {year}
  {2020}{\natexlab{a}})}\BibitemShut {NoStop}%
\bibitem [{\citenamefont {Haah}(2011)}]{Haah2011-ny}%
  \BibitemOpen
  \bibfield  {author} {\bibinfo {author} {\bibfnamefont {J.}~\bibnamefont
  {Haah}},\ }\bibfield  {title} {\bibinfo {title} {Local stabilizer codes in
  three dimensions without string logical operators},\ }\href@noop {}
  {\bibfield  {journal} {\bibinfo  {journal} {Phys. Rev. A}\ }\textbf {\bibinfo
  {volume} {83}},\ \bibinfo {pages} {042330} (\bibinfo {year}
  {2011})}\BibitemShut {NoStop}%
\bibitem [{\citenamefont {Vijay}\ and\ \citenamefont
  {Fu}(2017)}]{vijay2017generalization}%
  \BibitemOpen
  \bibfield  {author} {\bibinfo {author} {\bibfnamefont {S.}~\bibnamefont
  {Vijay}}\ and\ \bibinfo {author} {\bibfnamefont {L.}~\bibnamefont {Fu}},\
  }\bibfield  {title} {\bibinfo {title} {A generalization of non-abelian anyons
  in three dimensions},\ }\href@noop {} {\bibfield  {journal} {\bibinfo
  {journal} {arXiv preprint arXiv:1706.07070}\ } (\bibinfo {year}
  {2017})}\BibitemShut {NoStop}%
\bibitem [{\citenamefont {Chamon}(2005)}]{Chamon2005-fc}%
  \BibitemOpen
  \bibfield  {author} {\bibinfo {author} {\bibfnamefont {C.}~\bibnamefont
  {Chamon}},\ }\bibfield  {title} {\bibinfo {title} {Quantum glassiness in
  strongly correlated clean systems: an example of topological
  overprotection},\ }\href@noop {} {\bibfield  {journal} {\bibinfo  {journal}
  {Phys. Rev. Lett.}\ }\textbf {\bibinfo {volume} {94}},\ \bibinfo {pages}
  {040402} (\bibinfo {year} {2005})}\BibitemShut {NoStop}%
\bibitem [{\citenamefont {Pretko}(2017{\natexlab{a}})}]{pretko2017generalized}%
  \BibitemOpen
  \bibfield  {author} {\bibinfo {author} {\bibfnamefont {M.}~\bibnamefont
  {Pretko}},\ }\bibfield  {title} {\bibinfo {title} {Generalized
  electromagnetism of subdimensional particles: A spin liquid story},\
  }\href@noop {} {\bibfield  {journal} {\bibinfo  {journal} {Physical Review
  B}\ }\textbf {\bibinfo {volume} {96}},\ \bibinfo {pages} {035119} (\bibinfo
  {year} {2017}{\natexlab{a}})}\BibitemShut {NoStop}%
\bibitem [{\citenamefont
  {Pretko}(2017{\natexlab{b}})}]{pretko2017subdimensional}%
  \BibitemOpen
  \bibfield  {author} {\bibinfo {author} {\bibfnamefont {M.}~\bibnamefont
  {Pretko}},\ }\bibfield  {title} {\bibinfo {title} {Subdimensional particle
  structure of higher rank u (1) spin liquids},\ }\href@noop {} {\bibfield
  {journal} {\bibinfo  {journal} {Physical Review B}\ }\textbf {\bibinfo
  {volume} {95}},\ \bibinfo {pages} {115139} (\bibinfo {year}
  {2017}{\natexlab{b}})}\BibitemShut {NoStop}%
\bibitem [{\citenamefont {Vijay}\ \emph {et~al.}(2016)\citenamefont {Vijay},
  \citenamefont {Haah},\ and\ \citenamefont {Fu}}]{vijay2016fracton}%
  \BibitemOpen
  \bibfield  {author} {\bibinfo {author} {\bibfnamefont {S.}~\bibnamefont
  {Vijay}}, \bibinfo {author} {\bibfnamefont {J.}~\bibnamefont {Haah}},\ and\
  \bibinfo {author} {\bibfnamefont {L.}~\bibnamefont {Fu}},\ }\bibfield
  {title} {\bibinfo {title} {Fracton topological order, generalized lattice
  gauge theory, and duality},\ }\href@noop {} {\bibfield  {journal} {\bibinfo
  {journal} {Physical Review B}\ }\textbf {\bibinfo {volume} {94}},\ \bibinfo
  {pages} {235157} (\bibinfo {year} {2016})}\BibitemShut {NoStop}%
\bibitem [{\citenamefont {Yan}\ \emph {et~al.}(2019)\citenamefont {Yan},
  \citenamefont {Benton}, \citenamefont {Jaubert},\ and\ \citenamefont
  {Shannon}}]{yan2019rank}%
  \BibitemOpen
  \bibfield  {author} {\bibinfo {author} {\bibfnamefont {H.}~\bibnamefont
  {Yan}}, \bibinfo {author} {\bibfnamefont {O.}~\bibnamefont {Benton}},
  \bibinfo {author} {\bibfnamefont {L.~D.}\ \bibnamefont {Jaubert}},\ and\
  \bibinfo {author} {\bibfnamefont {N.}~\bibnamefont {Shannon}},\ }\bibfield
  {title} {\bibinfo {title} {Rank-2$ u (1) $ spin liquid on the breathing
  pyrochlore lattice},\ }\href@noop {} {\bibfield  {journal} {\bibinfo
  {journal} {arXiv preprint arXiv:1902.10934}\ } (\bibinfo {year}
  {2019})}\BibitemShut {NoStop}%
\bibitem [{\citenamefont {Xu}\ and\ \citenamefont {Fisher}(2007)}]{xu2007bond}%
  \BibitemOpen
  \bibfield  {author} {\bibinfo {author} {\bibfnamefont {C.}~\bibnamefont
  {Xu}}\ and\ \bibinfo {author} {\bibfnamefont {M.~P.~A.}\ \bibnamefont
  {Fisher}},\ }\bibfield  {title} {\bibinfo {title} {Bond algebraic liquid
  phase in strongly correlated multiflavor cold atom systems},\ }\href
  {https://doi.org/10.1103/PhysRevB.75.104428} {\bibfield  {journal} {\bibinfo
  {journal} {Phys. Rev. B}\ }\textbf {\bibinfo {volume} {75}},\ \bibinfo
  {pages} {104428} (\bibinfo {year} {2007})}\BibitemShut {NoStop}%
\bibitem [{\citenamefont {Rasmussen}\ and\ \citenamefont
  {Lu}(2018)}]{rasmussen2018intrinsically}%
  \BibitemOpen
  \bibfield  {author} {\bibinfo {author} {\bibfnamefont {A.}~\bibnamefont
  {Rasmussen}}\ and\ \bibinfo {author} {\bibfnamefont {Y.-M.}\ \bibnamefont
  {Lu}},\ }\bibfield  {title} {\bibinfo {title} {Intrinsically interacting
  topological crystalline insulators and superconductors},\ }\href@noop {}
  {\bibfield  {journal} {\bibinfo  {journal} {arXiv preprint arXiv:1810.12317}\
  } (\bibinfo {year} {2018})}\BibitemShut {NoStop}%
\bibitem [{\citenamefont {Prem}\ \emph {et~al.}(2018)\citenamefont {Prem},
  \citenamefont {Vijay}, \citenamefont {Chou}, \citenamefont {Pretko},\ and\
  \citenamefont {Nandkishore}}]{prem2018pinch}%
  \BibitemOpen
  \bibfield  {author} {\bibinfo {author} {\bibfnamefont {A.}~\bibnamefont
  {Prem}}, \bibinfo {author} {\bibfnamefont {S.}~\bibnamefont {Vijay}},
  \bibinfo {author} {\bibfnamefont {Y.-Z.}\ \bibnamefont {Chou}}, \bibinfo
  {author} {\bibfnamefont {M.}~\bibnamefont {Pretko}},\ and\ \bibinfo {author}
  {\bibfnamefont {R.~M.}\ \bibnamefont {Nandkishore}},\ }\bibfield  {title}
  {\bibinfo {title} {Pinch point singularities of tensor spin liquids},\
  }\href@noop {} {\bibfield  {journal} {\bibinfo  {journal} {arXiv preprint
  arXiv:1806.04148}\ } (\bibinfo {year} {2018})}\BibitemShut {NoStop}%
\bibitem [{\citenamefont {Gromov}\ \emph {et~al.}(2020)\citenamefont {Gromov},
  \citenamefont {Lucas},\ and\ \citenamefont
  {Nandkishore}}]{gromov2020fracton}%
  \BibitemOpen
  \bibfield  {author} {\bibinfo {author} {\bibfnamefont {A.}~\bibnamefont
  {Gromov}}, \bibinfo {author} {\bibfnamefont {A.}~\bibnamefont {Lucas}},\ and\
  \bibinfo {author} {\bibfnamefont {R.~M.}\ \bibnamefont {Nandkishore}},\
  }\bibfield  {title} {\bibinfo {title} {Fracton hydrodynamics},\ }\href@noop
  {} {\bibfield  {journal} {\bibinfo  {journal} {Physical Review Research}\
  }\textbf {\bibinfo {volume} {2}},\ \bibinfo {pages} {033124} (\bibinfo {year}
  {2020})}\BibitemShut {NoStop}%
\bibitem [{\citenamefont {Nandkishore}\ \emph {et~al.}(2021)\citenamefont
  {Nandkishore}, \citenamefont {Choi},\ and\ \citenamefont
  {Kim}}]{nandkishore2021spectroscopic}%
  \BibitemOpen
  \bibfield  {author} {\bibinfo {author} {\bibfnamefont {R.~M.}\ \bibnamefont
  {Nandkishore}}, \bibinfo {author} {\bibfnamefont {W.}~\bibnamefont {Choi}},\
  and\ \bibinfo {author} {\bibfnamefont {Y.~B.}\ \bibnamefont {Kim}},\
  }\bibfield  {title} {\bibinfo {title} {Spectroscopic fingerprints of gapped
  quantum spin liquids, both conventional and fractonic},\ }\href@noop {}
  {\bibfield  {journal} {\bibinfo  {journal} {Physical Review Research}\
  }\textbf {\bibinfo {volume} {3}},\ \bibinfo {pages} {013254} (\bibinfo {year}
  {2021})}\BibitemShut {NoStop}%
\bibitem [{\citenamefont {Shirley}\ \emph
  {et~al.}(2018{\natexlab{a}})\citenamefont {Shirley}, \citenamefont {Slagle},\
  and\ \citenamefont {Chen}}]{shirley2018fractional}%
  \BibitemOpen
  \bibfield  {author} {\bibinfo {author} {\bibfnamefont {W.}~\bibnamefont
  {Shirley}}, \bibinfo {author} {\bibfnamefont {K.}~\bibnamefont {Slagle}},\
  and\ \bibinfo {author} {\bibfnamefont {X.}~\bibnamefont {Chen}},\ }\bibfield
  {title} {\bibinfo {title} {Fractional excitations in foliated fracton
  phases},\ }\href@noop {} {\bibfield  {journal} {\bibinfo  {journal} {arXiv
  preprint arXiv:1806.08625}\ } (\bibinfo {year}
  {2018}{\natexlab{a}})}\BibitemShut {NoStop}%
\bibitem [{\citenamefont {Shirley}\ \emph
  {et~al.}(2018{\natexlab{b}})\citenamefont {Shirley}, \citenamefont {Slagle},\
  and\ \citenamefont {Chen}}]{shirley2018foliated}%
  \BibitemOpen
  \bibfield  {author} {\bibinfo {author} {\bibfnamefont {W.}~\bibnamefont
  {Shirley}}, \bibinfo {author} {\bibfnamefont {K.}~\bibnamefont {Slagle}},\
  and\ \bibinfo {author} {\bibfnamefont {X.}~\bibnamefont {Chen}},\ }\bibfield
  {title} {\bibinfo {title} {Foliated fracton order from gauging subsystem
  symmetries},\ }\href@noop {} {\bibfield  {journal} {\bibinfo  {journal}
  {arXiv preprint arXiv:1806.08679}\ } (\bibinfo {year}
  {2018}{\natexlab{b}})}\BibitemShut {NoStop}%
\bibitem [{\citenamefont {You}\ \emph {et~al.}(2020{\natexlab{b}})\citenamefont
  {You}, \citenamefont {Bibo},\ and\ \citenamefont {Pollmann}}]{you2020higher}%
  \BibitemOpen
  \bibfield  {author} {\bibinfo {author} {\bibfnamefont {Y.}~\bibnamefont
  {You}}, \bibinfo {author} {\bibfnamefont {J.}~\bibnamefont {Bibo}},\ and\
  \bibinfo {author} {\bibfnamefont {F.}~\bibnamefont {Pollmann}},\ }\bibfield
  {title} {\bibinfo {title} {Higher-order entanglement and many-body invariants
  for higher-order topological phases},\ }\href
  {https://doi.org/10.1103/PhysRevResearch.2.033192} {\bibfield  {journal}
  {\bibinfo  {journal} {Phys. Rev. Research}\ }\textbf {\bibinfo {volume}
  {2}},\ \bibinfo {pages} {033192} (\bibinfo {year}
  {2020}{\natexlab{b}})}\BibitemShut {NoStop}%
\bibitem [{\citenamefont {Seiberg}\ and\ \citenamefont
  {Shao}(2020)}]{seiberg2020exotic}%
  \BibitemOpen
  \bibfield  {author} {\bibinfo {author} {\bibfnamefont {N.}~\bibnamefont
  {Seiberg}}\ and\ \bibinfo {author} {\bibfnamefont {S.-H.}\ \bibnamefont
  {Shao}},\ }\bibfield  {title} {\bibinfo {title} {Exotic ${U}(1)$ symmetries,
  duality, and fractons in 3+1-dimensional quantum field theory},\ }\href
  {https://doi.org/10.21468/SciPostPhys.9.4.046} {\bibfield  {journal}
  {\bibinfo  {journal} {SciPost Phys.}\ }\textbf {\bibinfo {volume} {9}},\
  \bibinfo {pages} {46} (\bibinfo {year} {2020})}\BibitemShut {NoStop}%
\bibitem [{\citenamefont {Paramekanti}\ \emph {et~al.}(2002)\citenamefont
  {Paramekanti}, \citenamefont {Balents},\ and\ \citenamefont
  {Fisher}}]{paramekanti2002ring}%
  \BibitemOpen
  \bibfield  {author} {\bibinfo {author} {\bibfnamefont {A.}~\bibnamefont
  {Paramekanti}}, \bibinfo {author} {\bibfnamefont {L.}~\bibnamefont
  {Balents}},\ and\ \bibinfo {author} {\bibfnamefont {M.~P.~A.}\ \bibnamefont
  {Fisher}},\ }\bibfield  {title} {\bibinfo {title} {Ring exchange, the exciton
  bose liquid, and bosonization in two dimensions},\ }\href
  {https://doi.org/10.1103/PhysRevB.66.054526} {\bibfield  {journal} {\bibinfo
  {journal} {Phys. Rev. B}\ }\textbf {\bibinfo {volume} {66}},\ \bibinfo
  {pages} {054526} (\bibinfo {year} {2002})}\BibitemShut {NoStop}%
\bibitem [{\citenamefont {Karch}\ and\ \citenamefont
  {Raz}(2020)}]{karch2020reduced}%
  \BibitemOpen
  \bibfield  {author} {\bibinfo {author} {\bibfnamefont {A.}~\bibnamefont
  {Karch}}\ and\ \bibinfo {author} {\bibfnamefont {A.}~\bibnamefont {Raz}},\
  }\href@noop {} {\bibinfo {title} {Reduced conformal symmetry}} (\bibinfo
  {year} {2020}),\ \Eprint {https://arxiv.org/abs/2009.12308}
  {arXiv:2009.12308} \BibitemShut {NoStop}%
\bibitem [{\citenamefont {Gorantla}\ \emph {et~al.}(2021)\citenamefont
  {Gorantla}, \citenamefont {Lam}, \citenamefont {Seiberg},\ and\ \citenamefont
  {Shao}}]{gorantla2021modified}%
  \BibitemOpen
  \bibfield  {author} {\bibinfo {author} {\bibfnamefont {P.}~\bibnamefont
  {Gorantla}}, \bibinfo {author} {\bibfnamefont {H.~T.}\ \bibnamefont {Lam}},
  \bibinfo {author} {\bibfnamefont {N.}~\bibnamefont {Seiberg}},\ and\ \bibinfo
  {author} {\bibfnamefont {S.-H.}\ \bibnamefont {Shao}},\ }\href@noop {}
  {\bibinfo {title} {A modified villain formulation of fractons and other
  exotic theories}} (\bibinfo {year} {2021}),\ \Eprint
  {https://arxiv.org/abs/2103.01257} {arXiv:2103.01257} \BibitemShut {NoStop}%
\bibitem [{\citenamefont {You}\ \emph {et~al.}(2020{\natexlab{c}})\citenamefont
  {You}, \citenamefont {Bibo}, \citenamefont {Pollmann},\ and\ \citenamefont
  {Hughes}}]{you2020fracton}%
  \BibitemOpen
  \bibfield  {author} {\bibinfo {author} {\bibfnamefont {Y.}~\bibnamefont
  {You}}, \bibinfo {author} {\bibfnamefont {J.}~\bibnamefont {Bibo}}, \bibinfo
  {author} {\bibfnamefont {F.}~\bibnamefont {Pollmann}},\ and\ \bibinfo
  {author} {\bibfnamefont {T.~L.}\ \bibnamefont {Hughes}},\ }\href@noop {}
  {\bibinfo {title} {Fracton critical point in higher-order topological phase
  transition}} (\bibinfo {year} {2020}{\natexlab{c}}),\ \Eprint
  {https://arxiv.org/abs/2008.01746} {arXiv:2008.01746} \BibitemShut {NoStop}%
\bibitem [{\citenamefont {You}\ \emph {et~al.}(2021)\citenamefont {You},
  \citenamefont {Bibo}, \citenamefont {Hughes},\ and\ \citenamefont
  {Pollmann}}]{you2021fractonic}%
  \BibitemOpen
  \bibfield  {author} {\bibinfo {author} {\bibfnamefont {Y.}~\bibnamefont
  {You}}, \bibinfo {author} {\bibfnamefont {J.}~\bibnamefont {Bibo}}, \bibinfo
  {author} {\bibfnamefont {T.~L.}\ \bibnamefont {Hughes}},\ and\ \bibinfo
  {author} {\bibfnamefont {F.}~\bibnamefont {Pollmann}},\ }\bibfield  {title}
  {\bibinfo {title} {Fractonic critical point proximate to a higher-order
  topological insulator: How uv blend with ir?},\ }\href@noop {} {\bibfield
  {journal} {\bibinfo  {journal} {arXiv preprint arXiv:2101.01724}\ } (\bibinfo
  {year} {2021})}\BibitemShut {NoStop}%
\bibitem [{\citenamefont {Gromov}(2019)}]{gromov2019towards}%
  \BibitemOpen
  \bibfield  {author} {\bibinfo {author} {\bibfnamefont {A.}~\bibnamefont
  {Gromov}},\ }\bibfield  {title} {\bibinfo {title} {Towards classification of
  fracton phases: The multipole algebra},\ }\href
  {https://doi.org/10.1103/PhysRevX.9.031035} {\bibfield  {journal} {\bibinfo
  {journal} {Phys. Rev. X}\ }\textbf {\bibinfo {volume} {9}},\ \bibinfo {pages}
  {031035} (\bibinfo {year} {2019})}\BibitemShut {NoStop}%
\bibitem [{\citenamefont {Benton}\ and\ \citenamefont
  {Moessner}(2021)}]{benton2021topological}%
  \BibitemOpen
  \bibfield  {author} {\bibinfo {author} {\bibfnamefont {O.}~\bibnamefont
  {Benton}}\ and\ \bibinfo {author} {\bibfnamefont {R.}~\bibnamefont
  {Moessner}},\ }\bibfield  {title} {\bibinfo {title} {Topological route to new
  and unusual coulomb spin liquids},\ }\href@noop {} {\bibfield  {journal}
  {\bibinfo  {journal} {arXiv preprint arXiv:2103.10817}\ } (\bibinfo {year}
  {2021})}\BibitemShut {NoStop}%
\bibitem [{\citenamefont {Landau}\ \emph {et~al.}(2009)\citenamefont {Landau},
  \citenamefont {M.},\ and\ \citenamefont {P.}}]{landau}%
  \BibitemOpen
  \bibfield  {author} {\bibinfo {author} {\bibfnamefont {L.~D.}\ \bibnamefont
  {Landau}}, \bibinfo {author} {\bibfnamefont {L.~E.}\ \bibnamefont {M.}},\
  and\ \bibinfo {author} {\bibfnamefont {P.~L.}\ \bibnamefont {P.}},\
  }\href@noop {} {\emph {\bibinfo {title} {Statistical physics}}}\ (\bibinfo
  {publisher} {Butterworth-Heinemann},\ \bibinfo {year} {2009})\BibitemShut
  {NoStop}%
\bibitem [{\citenamefont {Wilson}(1983)}]{wilson1983renormalization}%
  \BibitemOpen
  \bibfield  {author} {\bibinfo {author} {\bibfnamefont {K.~G.}\ \bibnamefont
  {Wilson}},\ }\bibfield  {title} {\bibinfo {title} {The renormalization group
  and critical phenomena},\ }\href {https://doi.org/10.1103/RevModPhys.55.583}
  {\bibfield  {journal} {\bibinfo  {journal} {Rev. Mod. Phys.}\ }\textbf
  {\bibinfo {volume} {55}},\ \bibinfo {pages} {583} (\bibinfo {year}
  {1983})}\BibitemShut {NoStop}%
\bibitem [{\citenamefont {Fisher}(1983)}]{fisher1983scaling}%
  \BibitemOpen
  \bibfield  {author} {\bibinfo {author} {\bibfnamefont {M.~E.}\ \bibnamefont
  {Fisher}},\ }\bibfield  {title} {\bibinfo {title} {Scaling, universality and
  renormalization group theory},\ }in\ \href@noop {} {\emph {\bibinfo
  {booktitle} {Critical phenomena}}}\ (\bibinfo  {publisher} {Springer},\
  \bibinfo {year} {1983})\ pp.\ \bibinfo {pages} {1--139}\BibitemShut {NoStop}%
\bibitem [{\citenamefont {Fisher}(1998)}]{fisher1998renormalization}%
  \BibitemOpen
  \bibfield  {author} {\bibinfo {author} {\bibfnamefont {M.~E.}\ \bibnamefont
  {Fisher}},\ }\bibfield  {title} {\bibinfo {title} {Renormalization group
  theory: Its basis and formulation in statistical physics},\ }\href
  {https://doi.org/10.1103/RevModPhys.70.653} {\bibfield  {journal} {\bibinfo
  {journal} {Rev. Mod. Phys.}\ }\textbf {\bibinfo {volume} {70}},\ \bibinfo
  {pages} {653} (\bibinfo {year} {1998})}\BibitemShut {NoStop}%
\bibitem [{\citenamefont {You}\ \emph {et~al.}(2019)\citenamefont {You},
  \citenamefont {Devakul}, \citenamefont {Sondhi},\ and\ \citenamefont
  {Burnell}}]{you2019fractonic}%
  \BibitemOpen
  \bibfield  {author} {\bibinfo {author} {\bibfnamefont {Y.}~\bibnamefont
  {You}}, \bibinfo {author} {\bibfnamefont {T.}~\bibnamefont {Devakul}},
  \bibinfo {author} {\bibfnamefont {S.}~\bibnamefont {Sondhi}},\ and\ \bibinfo
  {author} {\bibfnamefont {F.}~\bibnamefont {Burnell}},\ }\bibfield  {title}
  {\bibinfo {title} {Fractonic chern-simons and bf theories},\ }\href@noop {}
  {\bibfield  {journal} {\bibinfo  {journal} {arXiv preprint arXiv:1904.11530}\
  } (\bibinfo {year} {2019})}\BibitemShut {NoStop}%
\bibitem [{Note1()}]{Note1}%
  \BibitemOpen
  \bibinfo {note} {Such defect illustrates a configuration where a vortex line
  oriented along y(x) terminated at a point.}\BibitemShut {Stop}%
\bibitem [{\citenamefont {Kosterlitz}\ and\ \citenamefont
  {Thouless}(1973)}]{kosterlitz1973ordering}%
  \BibitemOpen
  \bibfield  {author} {\bibinfo {author} {\bibfnamefont {J.~M.}\ \bibnamefont
  {Kosterlitz}}\ and\ \bibinfo {author} {\bibfnamefont {D.~J.}\ \bibnamefont
  {Thouless}},\ }\bibfield  {title} {\bibinfo {title} {Ordering, metastability
  and phase transitions in two-dimensional systems},\ }\href@noop {} {\bibfield
   {journal} {\bibinfo  {journal} {Journal of Physics C: Solid State Physics}\
  }\textbf {\bibinfo {volume} {6}},\ \bibinfo {pages} {1181} (\bibinfo {year}
  {1973})}\BibitemShut {NoStop}%
\bibitem [{Note2()}]{Note2}%
  \BibitemOpen
  \bibinfo {note} {In the quantum version of this liquid theory\cite
  {gorantla2021modified}, the emergent $U(1)\times U(1)$ subsystem symmetry is
  responsible for the mixed 't-Hooft anomaly.}\BibitemShut {Stop}%
\bibitem [{\citenamefont {Bulmash}\ and\ \citenamefont
  {Barkeshli}(2018)}]{bulmash2018generalized}%
  \BibitemOpen
  \bibfield  {author} {\bibinfo {author} {\bibfnamefont {D.}~\bibnamefont
  {Bulmash}}\ and\ \bibinfo {author} {\bibfnamefont {M.}~\bibnamefont
  {Barkeshli}},\ }\href@noop {} {\bibinfo {title} {Generalized ${U}(1)$ gauge
  field theories and fractal dynamics}} (\bibinfo {year} {2018}),\ \Eprint
  {https://arxiv.org/abs/1806.01855} {arXiv:1806.01855} \BibitemShut {NoStop}%
\bibitem [{\citenamefont {Newman}\ and\ \citenamefont
  {Moore}(1999)}]{newman_1}%
  \BibitemOpen
  \bibfield  {author} {\bibinfo {author} {\bibfnamefont {M.~E.~J.}\
  \bibnamefont {Newman}}\ and\ \bibinfo {author} {\bibfnamefont
  {C.}~\bibnamefont {Moore}},\ }\bibfield  {title} {\bibinfo {title} {Glassy
  dynamics and aging in an exactly solvable spin model},\ }\href
  {https://doi.org/10.1103/PhysRevE.60.5068} {\bibfield  {journal} {\bibinfo
  {journal} {Phys. Rev. E}\ }\textbf {\bibinfo {volume} {60}},\ \bibinfo
  {pages} {5068} (\bibinfo {year} {1999})}\BibitemShut {NoStop}%
\bibitem [{\citenamefont {Garrahan}\ and\ \citenamefont
  {Newman}(2000)}]{newman_2}%
  \BibitemOpen
  \bibfield  {author} {\bibinfo {author} {\bibfnamefont {J.~P.}\ \bibnamefont
  {Garrahan}}\ and\ \bibinfo {author} {\bibfnamefont {M.~E.~J.}\ \bibnamefont
  {Newman}},\ }\bibfield  {title} {\bibinfo {title} {Glassiness and constrained
  dynamics of a short-range nondisordered spin model},\ }\href
  {https://doi.org/10.1103/PhysRevE.62.7670} {\bibfield  {journal} {\bibinfo
  {journal} {Phys. Rev. E}\ }\textbf {\bibinfo {volume} {62}},\ \bibinfo
  {pages} {7670} (\bibinfo {year} {2000})}\BibitemShut {NoStop}%
\bibitem [{\citenamefont {Zhou}\ \emph {et~al.}(2021)\citenamefont {Zhou},
  \citenamefont {Zhang}, \citenamefont {Pollmann},\ and\ \citenamefont
  {You}}]{zhou2021fractal}%
  \BibitemOpen
  \bibfield  {author} {\bibinfo {author} {\bibfnamefont {Z.}~\bibnamefont
  {Zhou}}, \bibinfo {author} {\bibfnamefont {X.-F.}\ \bibnamefont {Zhang}},
  \bibinfo {author} {\bibfnamefont {F.}~\bibnamefont {Pollmann}},\ and\
  \bibinfo {author} {\bibfnamefont {Y.}~\bibnamefont {You}},\ }\bibfield
  {title} {\bibinfo {title} {Fractal quantum phase transitions: Critical
  phenomena beyond renormalization},\ }\href@noop {} {\bibfield  {journal}
  {\bibinfo  {journal} {arXiv preprint arXiv:2105.05851}\ } (\bibinfo {year}
  {2021})}\BibitemShut {NoStop}%
\end{thebibliography}
\end{document}